\documentclass[twocolumn]{aa}

\usepackage{aas_macros}
\usepackage{natbib}
\usepackage{scalerel,stackengine,amsmath}
\usepackage{siunitx}
\usepackage{amssymb}
\usepackage{grffile}
\usepackage{graphicx}   % need for figures
\usepackage{verbatim}   % useful for program listings
\usepackage{color}      % use if color is used in text
\usepackage{subcaption}  % use for side-by-side figures
\usepackage{hyperref}   % use for hypertext links, including those to external documents and URLs
\usepackage{tikz}
\usepackage[export]{adjustbox}
\usetikzlibrary{shapes,arrows}
\newcommand\equalhat{\mathrel{\stackon[1.5pt]{=}{\stretchto{%
    \scalerel*[\widthof{=}]{\wedge}{\rule{1ex}{3ex}}}{0.5ex}}}}

\tikzstyle{block} = [rectangle,thick, draw=black!100, %fill=blue!30, 
    text width=1.2em, text centered, rounded corners, minimum height=1em,minimum width = 0.6em]
\tikzstyle{block2} = [rectangle,thick, draw=black!100, %fill=blue!30, 
    text width=2.4em, text centered, rounded corners, minimum height=1em,minimum width = 1.2em]
\tikzstyle{line} = [draw,ultra thick, -latex']
\tikzstyle{curved line} = [draw, bend right=45,ultra thick, -latex']
\tikzstyle{dashed line} = [draw,ultra thick, -latex',dashed]
\tikzstyle{cloud} = [thick, ellipse, draw=black!100, %fill=green!30, 
    node distance=0.6cm, minimum width=2em,
    minimum height=1.2em] 

\bibpunct{(}{)}{;}{a}{}{,} % to follow the A&A style

\begin{document}

\title{The Galactic Faraday depth sky revisited}
\author{Sebastian Hutschenreuter \and Torsten A. En{\ss}lin}
\institute{Max Planck Institute for Astrophysics, Karl-Schwarzschildstr.1, 85741 Garching, Germany \and Ludwig-Maximilians-Universit\"at M\"unchen, Geschwister-Scholl-Platz 1, 80539 Munich, Germany}

\abstract{
The Galactic Faraday depth sky is a tracer for both the Galactic magnetic field and the thermal electron distribution. It has been previously reconstructed from polarimetric measurements of extra-galactic point sources. Here, we improve on these works by using an updated inference algorithm as well as by taking into account the free-free emission measure map from the Planck survey. In the future, the data situation will improve drastically with the next generation Faraday rotation measurements from SKA and its pathfinders. Anticipating this, the aim of this paper is to update the map reconstruction method with the latest development in imaging based on information field theory. We demonstrate the validity of the new algorithm by applying it to the \citet{OLD_FARADAY_2012A&A...542A..93O} data compilation and compare our results to the previous map.\\
Despite using exactly the previous data set, a number of novel findings are made: A non-parametric reconstruction of an overall amplitude field resembles the free-free emission measure map of the Galaxy. Folding this free-free map into the analysis allows for more detailed predictions.
The joint inference enables us to identify regions with deviations from the assumed correlations between the free-free and Faraday data, thereby pointing us to Galactic structures with distinguishably different physics. We e.g. find evidence for an alignment of the magnetic field within the line of sights along both directions of the Orion arm.
}
\maketitle

\section{Introduction}
\label{sec:intro}

\begin{figure*}
\centering
\begin{subfigure}{\textwidth}
\includegraphics[width=\textwidth]{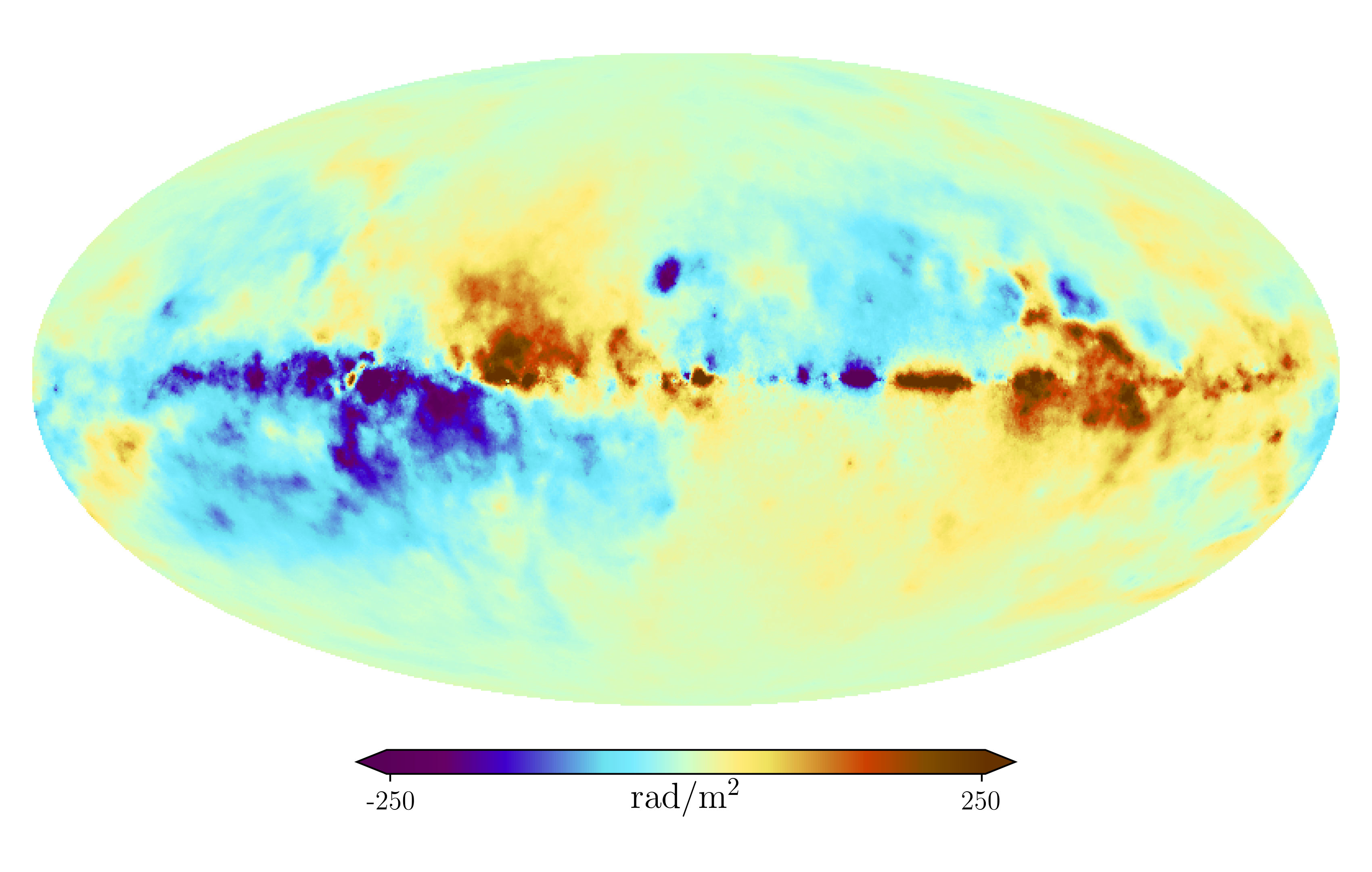}
\caption{\label{fig:rev_faraday_mean}}
\end{subfigure}
\begin{subfigure}{0.49\textwidth}
\includegraphics[width=\textwidth]{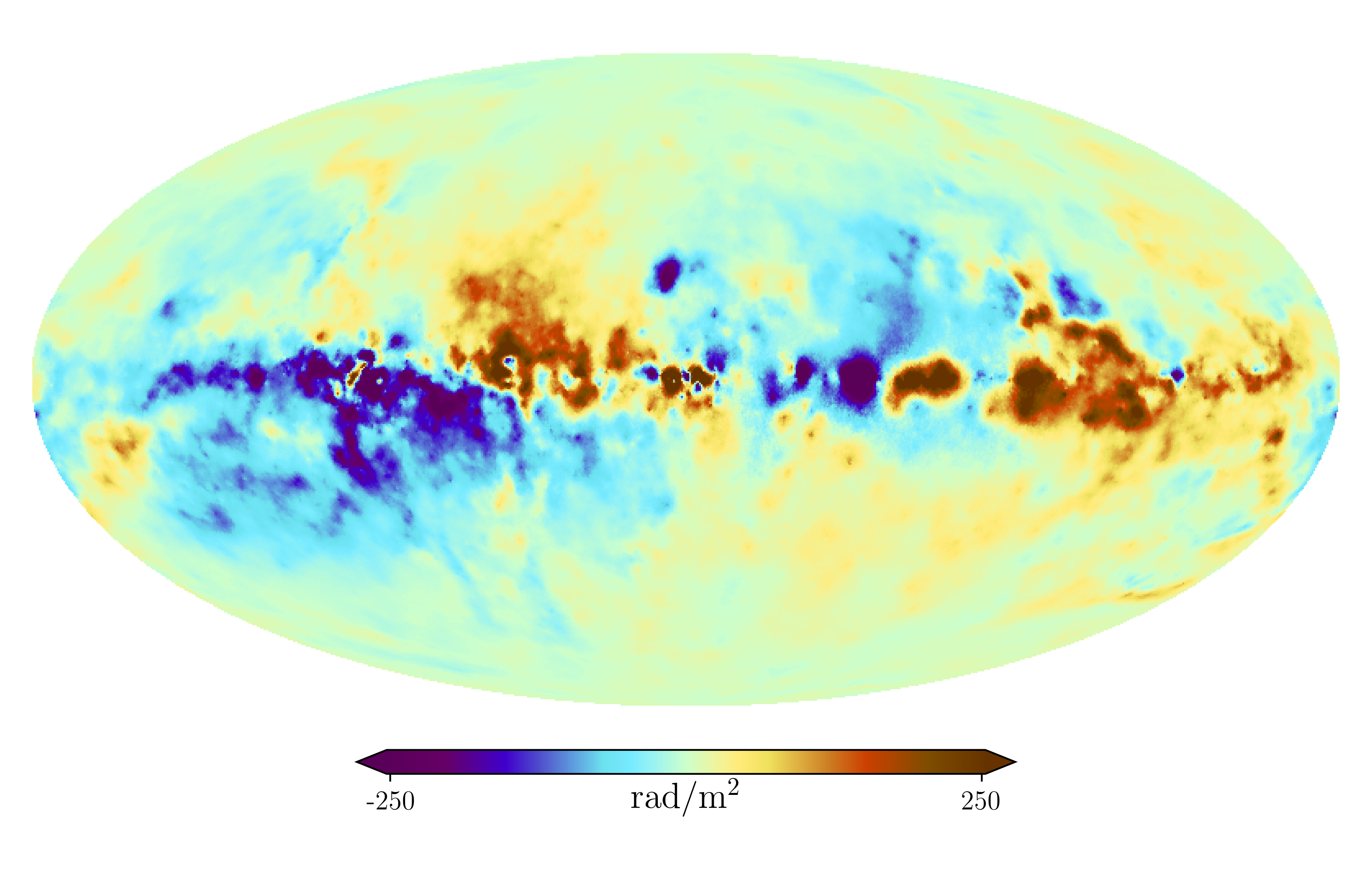}
\caption{\label{fig:ini_faraday_mean} }
\end{subfigure}
\begin{subfigure}{0.49\textwidth}
\centering
\includegraphics[width=\textwidth]{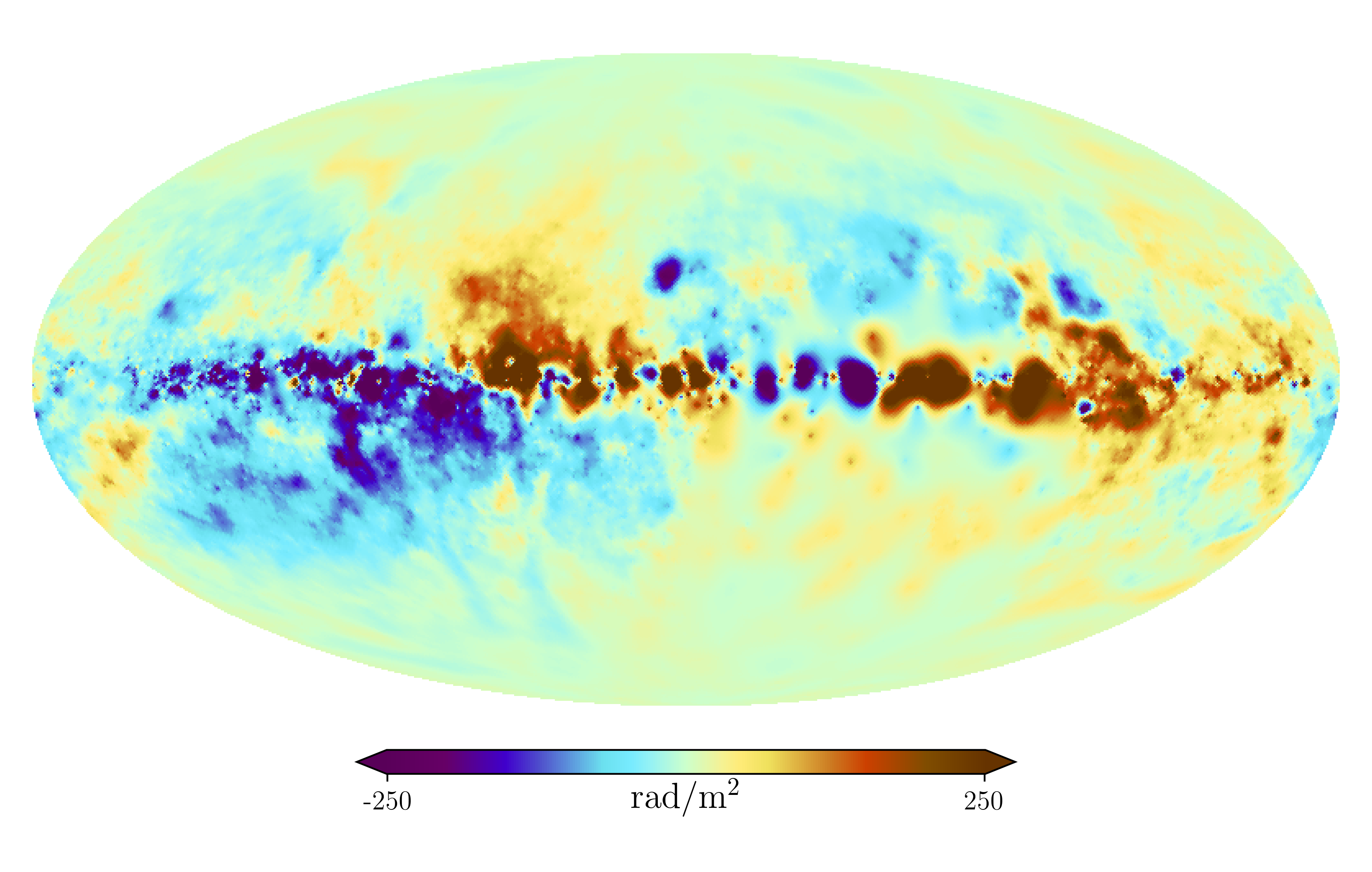}
\caption{\label{fig:pre_faraday_mean}}
\end{subfigure}
\caption{\label{fig:faraday_mean} Reconstructions of the Faraday sky. Figure (a) shows our main result. This map uses the free-free EM map as a proxy for the Faraday amplitude. The model furthermore contains fields that translate the free-free map into a Faraday amplitude, thereby balancing between effects that support and such that disturb a direct relation of these two quantities. Figure (b) contains the posterior mean of the initially revised reconstruction. Here we did not use the additional free-free data. Figure (c) shows the posterior mean of the previous reconstruction \citep{EXTRAGAL_2015A&A...575A.118O}. We show the differences of the three maps in Fig. \ref{fig:diff}.
}
\end{figure*}

The Faraday rotation effect is one of the primary sources of information on astrophysical and cosmological magnetic fields. This includes the fields of planets, stars, other galaxies and galaxy clusters, as well as more curious objects such as radio jets, lobes, and relics.
In particular, the study of the Faraday rotation induced by the Milky Way magnetic field is of twofold importance. 
First of all it is an interesting research topic on its own, due its connection to the formation and structure of our home galaxy, but furthermore it also constitutes a non-negligible foreground component. All polarized light stemming from cosmological sources passes through the Galaxy and interacts with the Galactic magnetic field, which affects its polarization direction. Accurate and highly resolved Galactic foreground templates are therefore a necessary condition for any precision measurement of extra-galactic magnetic fields via Faraday rotation.\\
Multiple efforts have been made to to map the Galactic Faraday sky, quite a few of them already to remarkable accuracy, e.g. in  \citet{Frick2001MNRAS.325..649F, JohnstonHollit2004mim..proc...13J, Dineen2005MNRAS.362..403D, Xu2006ApJ...637...19X, OLD_FARADAY_2012A&A...542A..93O} and also in \citet{EXTRAGAL_2015A&A...575A.118O}. 
Especially the approach in \citet{OLD_FARADAY_2012A&A...542A..93O} is important to us, as we will rely on the same theoretical framework for our inference algorithms. \\
Our goal in this work is to further sharpen our knowledge on the Faraday sky by exploiting correlations to bremsstrahlung measured from electron proton interaction in the interstellar medium, commonly known as the Galactic free-free emission. This is well motivated by observation as well as physical considerations, as we will outline later on. In a more general context, we hope to demonstrate an interesting test case of multi-messenger astronomy, and show that the synthesis of long existing data sets still can yield undiscovered information, under the condition that our a priori physical knowledge is employed and the data is treated in a consistent way. Our main result, the revised Faraday sky including the free-free data is shown in Fig. \ref{fig:rev_faraday_mean} and the revised map without free-free data in Fig. \ref{fig:ini_faraday_mean}. The results of our inference are publicly available \footnote{\url{https://wwwmpa.mpa-garching.mpg.de/~ensslin/research/data/faraday_revisited.html}}.\\
The paper is structured as follows. In Sec. \ref{sec:improving_the_inference_of_the_faraday_sky} we develop our updated reconstruction method. This involves an amplitude field aiming at representing the overall Galactic structure and a sign field representing magnetic reversals and smaller scaled structures. It turns out that the reconstructed amplitude field resembles the free-free emission measure (EM) of the Galaxy, which is physically plausible. In order to take advantage of this, we extend the method to fold in the measurements of the free-free sky in Sec. \ref{sec:including_free_free}. We conclude in Sec. \ref{sec:conclusion}. Mathematical details of the method can be found in the Appendix. 

\begin{figure}
\begin{subfigure}{0.99\linewidth}
\centering
\includegraphics[width=1.0\textwidth]{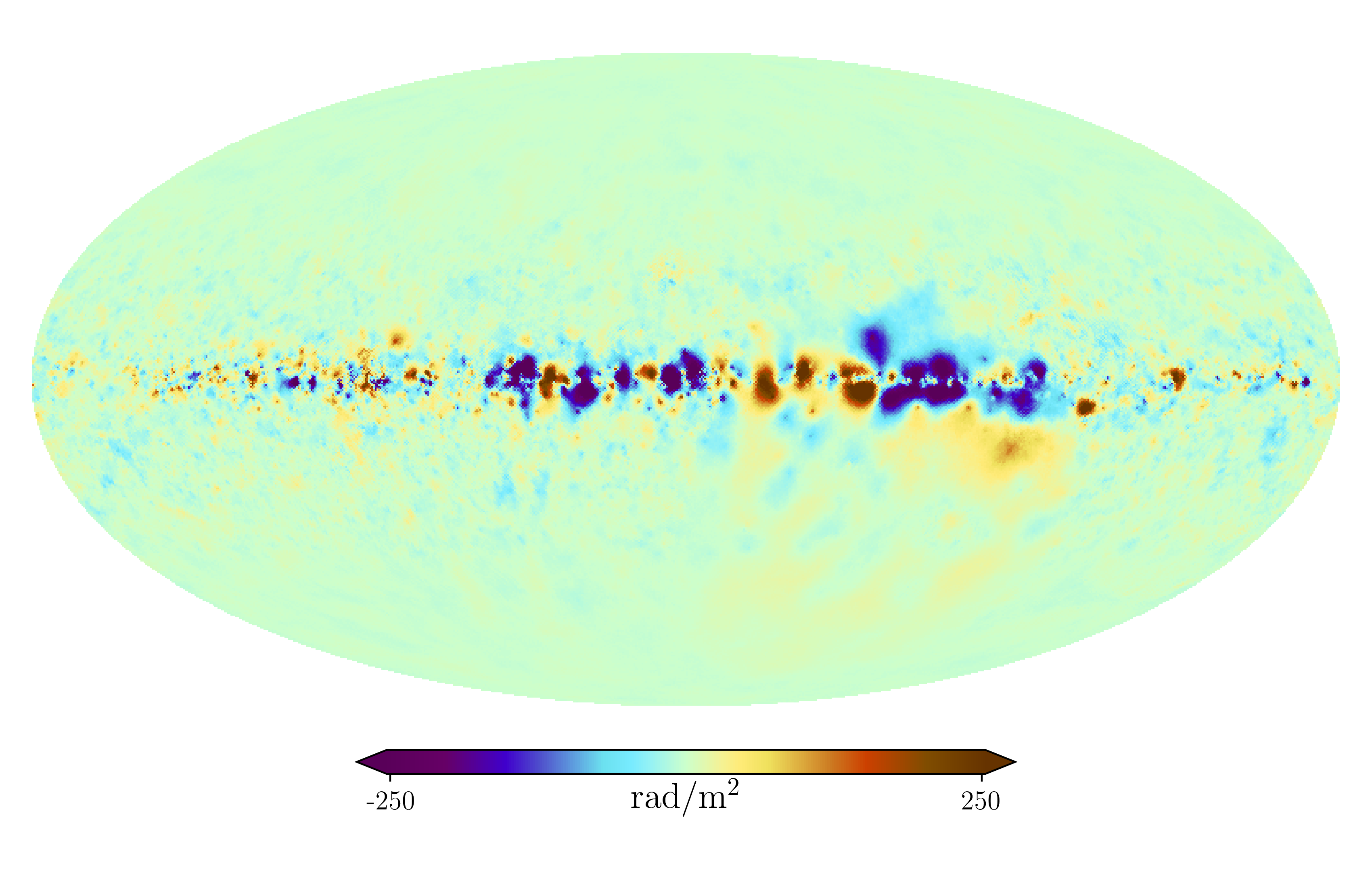}
\caption{\label{fig:diff_pre_rev}}
\end{subfigure}
\begin{subfigure}{0.99\linewidth}
\centering
\includegraphics[width=1.0\textwidth]{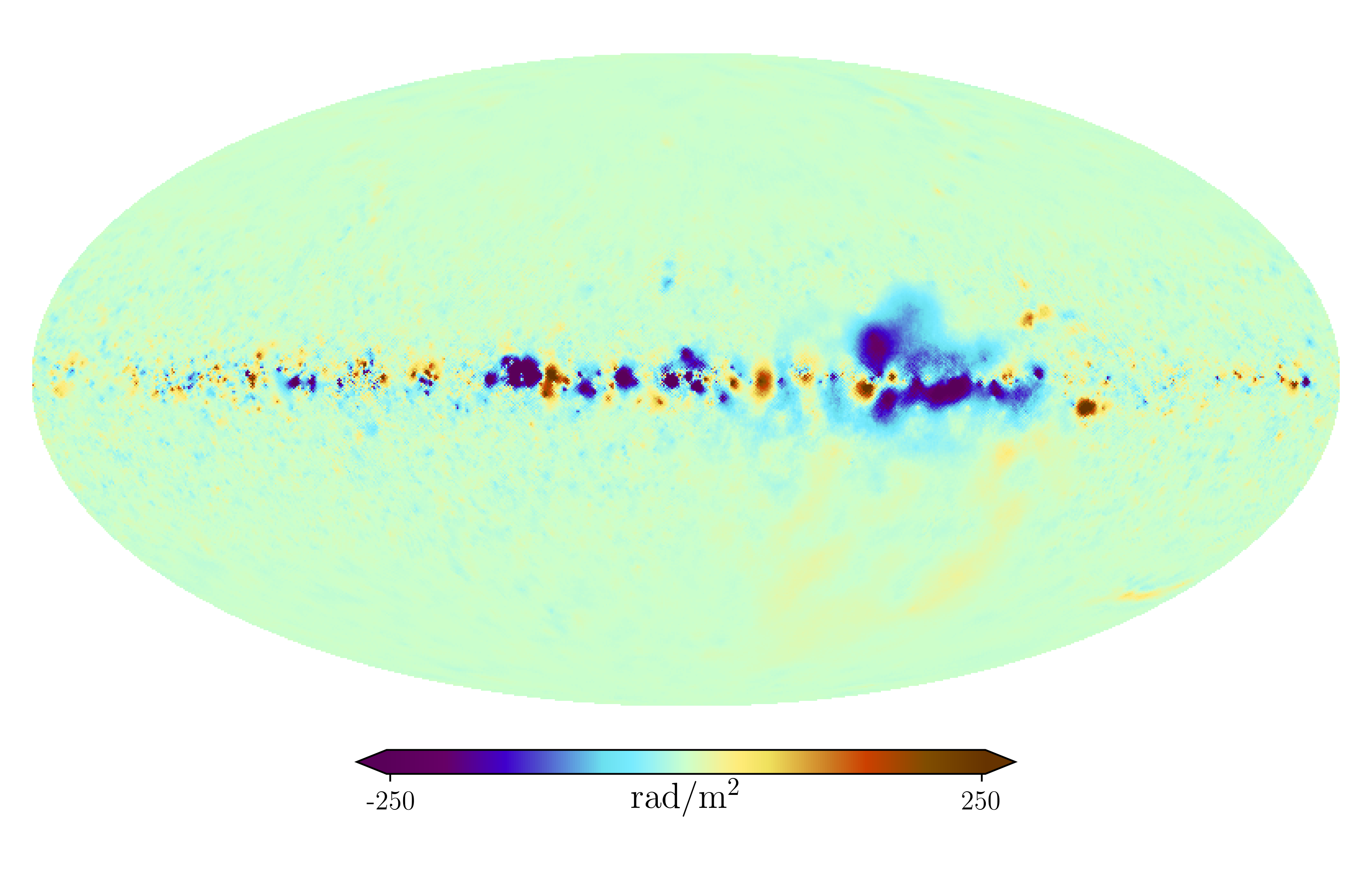}
\caption{\label{fig:diff_pre_ini}}
\end{subfigure}
\begin{subfigure}{0.99\linewidth}
\centering
\includegraphics[width=1.0\textwidth]{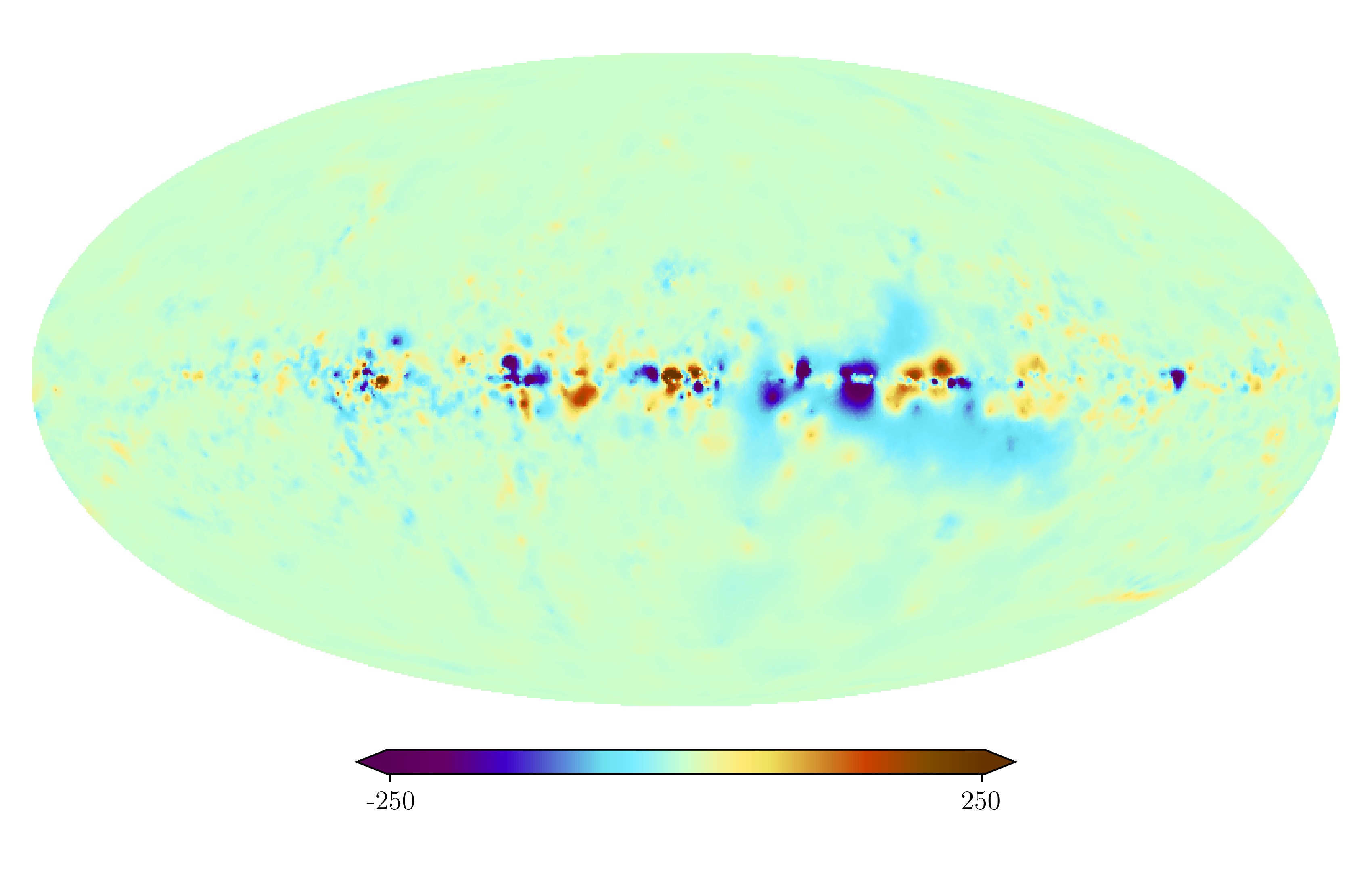}
\caption{\label{fig:diff_rev_ini}}
\end{subfigure}
\caption{\label{fig:diff} Differences of the reconstructions of the Faraday sky shown in Fig. \ref{fig:faraday_mean}. Figure (a) shows the difference between the previous map and our revised reconstruction including the free free data. Figure (b) shows the difference between the previous map and our initially revised reconstruction. Figure (c) finally shows the difference between the Faraday sky resulting from our revised reconstruction including the free free data and the initially revised sky.} 
\end{figure}

\section{Improving the Inference of the Faraday sky}
\label{sec:improving_the_inference_of_the_faraday_sky}

\subsection{The physics}
\label{subsec:the_physics_faraday}

The rotation angle $\Delta_\lambda$ of the polarization plane of linearly polarized light due to Faraday rotation can be written in most astrophysical settings as \citep[see e.g.][]{OLD_FARADAY_2012A&A...542A..93O, EXTRAGAL_2015A&A...575A.118O}
\begin{equation}
\label{eq:faraday_angle}
\Delta_\lambda = \phi\left(n_{\mathrm{th}}, B_{\mathrm{LOS}}\right)\lambda^2,
\end{equation}
where $\lambda$ is the wavelength at which the light is observed. The Faraday  
depth $\phi$ depends on the thermal electron density  $n_{\mathrm{th}}$  and the line of sight (LOS) component of the magnetic field $B_{\mathrm{LOS}}$. More specifically for a polarized extra-galactic source shining through the interstellar medium, the Galactic part of the Faraday depth $\phi_{\mathrm{gal}}$ can be written as \citep{OLD_FARADAY_2012A&A...542A..93O}

\begin{equation}
\label{eq:faraday_physics}
\phi_{\mathrm{gal}} = \frac{e^3}{2\pi m_e^2 c^4}\int_{\mathrm{LOS}} dl\, n_{\mathrm{th}} B_{\mathrm{gal, LOS}},
\end{equation}
where $B_{\mathrm{gal}}$ is the galactic magnetic field, $e$ is the electron charge, $m_e$ is the electron mass and $c$ the speed of light. 
Unfortunately, contributions to the observed rotation angle not only come from the Galactic magnetic field but also contain extra-galactic, potentially source internal contributions. 
Similar to the approach taken by \citet{OLD_FARADAY_2012A&A...542A..93O}, it will be helpful to separate the Faraday depth signal into
\begin{equation}
\label{eq:faraday_separation}
\phi = \phi_{\mathrm{gal}} +  \phi_{\mathrm{etc}}.  
\end{equation}

The terms $\phi_{\mathrm{gal}}$ and $\phi_{\mathrm{etc}} $ describe the aforementioned Galactic and other e.g. extra-galactic, source intrinsic, or ionospheric contributions, respectively. Our goal in this work will be the estimation of $\phi_{\mathrm{gal}}$ as a correlated field on the sky. This implies the necessity for an at least stochastic estimation of the other contributions. The connection of $\phi$ to the observed data $d_\phi$ is the topic of the next section.

\subsection{The data}
\label{subsec:the_data_faraday}

\begin{figure}
\begin{subfigure}{\linewidth}
\includegraphics[width=1.0\textwidth, left]{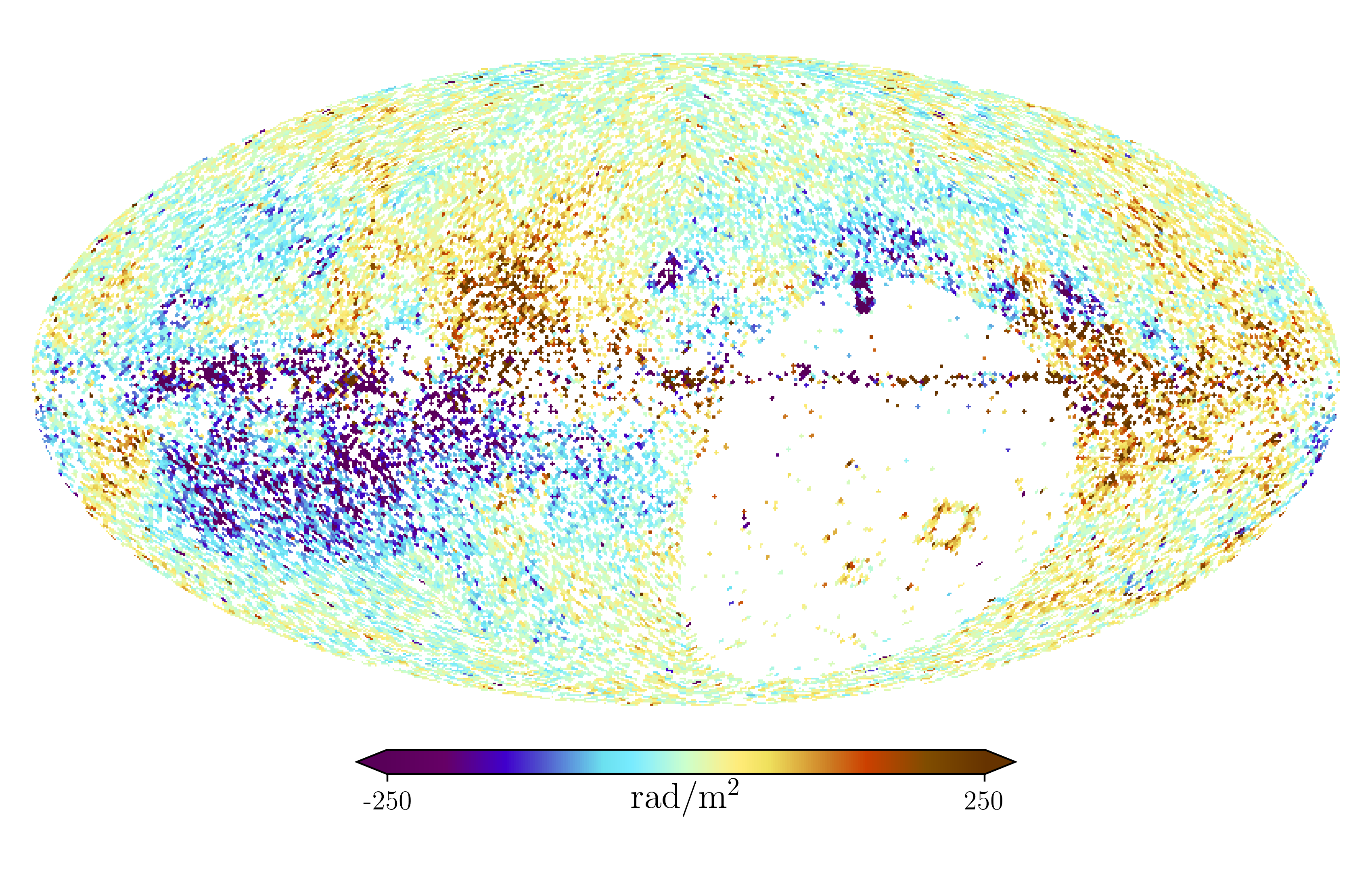}
\caption{\label{fig:faraday_data} }
\end{subfigure}
\begin{subfigure}{0.49\textwidth}
\centering%
\includegraphics[width=1.0\textwidth]{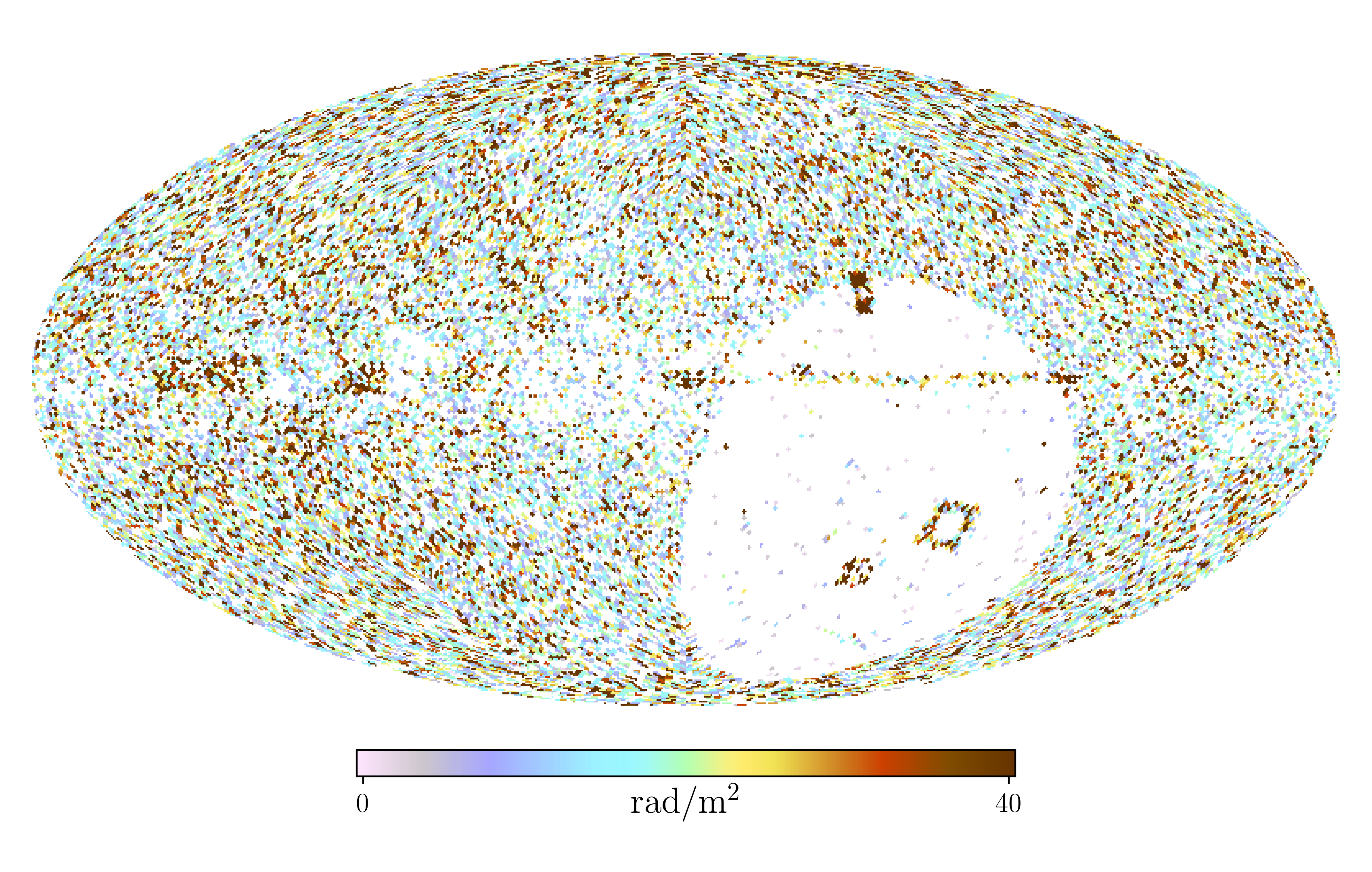}
\caption{\label{fig:faraday_rms_data}}
\end{subfigure}
\caption{\label{fig:data} Sky map of the RM data set (figure (a)) used in this work and the corresponding uncertainties in figure (b). The resolution of the sky pixelisation is lower in this image than in the reconstruction for illustrational purposes. We note that the region corresponding to the terrestrial south pole is only weakly constrained by data, except for the Galactic disk.}
\end{figure}

A measurement of the Galactic Faraday depth can be modeled via

\begin{equation}
\label{eq:faraday_data}
d_\phi =  \mathcal{R}_\phi \phi  + n_\phi,  
\end{equation}
where we ignore for the moment the potential non-galactic contributions mentioned in Eq. \eqref{eq:faraday_separation}. The operator $ \mathcal{R}_\phi$ is called the response and describes the (noiseless) measurement process translating the signal into data space.
In this case, the evaluation of the response at the location of source $i$ is sufficiently described via
\begin{equation}
\label{eq:faraday_response}
\left[\mathcal{R} \phi \right]_i  = \int_{S^2} d\widehat{e}\,\, \phi\left(\widehat{e}\right)  \delta(\widehat{e}-\widehat{e}_i)  
\end{equation} 
with $\widehat{e}$ being the radial unit vector on the 2-sphere.
The quantity $n_\phi$ in Eq. \eqref{eq:faraday_data} is the measurement noise, which in the easiest case can be assumed to be independent and Gaussian distributed, according to

\begin{equation}
\label{eq:faraday_noise_pdf}
\mathcal{P}\left(n_\phi\right) = \mathcal{G}\left(n_\phi, N_{\phi}\right) = \frac{1}{\sqrt{\vert 2\pi N_{\phi}\vert}}e^{-\frac{1}{2} n_\phi N_{\phi}^{-1} n_{\phi}^\dagger},
\end{equation}
with zero mean and known noise covariance $N_{\phi}$. The noise will play an important role for our estimation of non-galactic contributions to $\phi$, which will be discussed in more detail in Sec. \ref{subsubsec:the_noise_model_faraday}. For the moment we assume $N_{\phi}$ to be simply given by the observational measurement uncertainties, therefore being a diagonal matrix in data space. \\
As we are conducting a Bayesian inference, our main task is the evaluation of the posterior probability distribution. An integral part of this distribution is the likelihood, which in our case is best described by the following Gaussian:
\begin{equation}
\label{eq:faraday_likelihood}
\mathcal{P}\left(d_\phi|\phi\right) = \mathcal{G}\left(d_\phi - \mathcal{R}_\phi\phi,N_\phi\right)
\end{equation}
The likelihood will be further determined by the the specific modeling of $\phi$, which we will present in Sec. \ref{fig:sky_model_faraday}. There we will also specify the different prior terms.\\
The data set used in this work is exactly the same compilation of Faraday rotation measures as used in \citet{OLD_FARADAY_2012A&A...542A..93O} and is mostly publicly accessible \footnote{\url{https://wwwmpa.mpa-garching.mpg.de/ift/faraday/}}. This compilation contains rotation measures, corresponding standard deviations and source positions for 41632 sources \footnote{\citep{CAT_01_2010A&A...513A..30B, CAT_02_Broten1988, CAT_03_Brown_2003, CAT_04_Taylor_2003, CAT_05_2007ApJ...663..258B, CAT_06_2006ApJS..167..230H, CAT_07_2005ApJS..158..178M, CAT_08_2001ApJ...547L.111C, CAT_09_2004JKAS...37..337C, CAT_10_1992ApJ...386..143C, CAT_11_2009ApJ...707..114F, CAT_12_2011ApJ...740...17F, CAT_13_2001ApJ...549..959G, CAT_14, CAT_15_2009A&A...503..409H, CAT_16_2007A&A...461..455B, CAT_17_1989ApJ...347..144H, CAT_18, CAT_19_Johnston-HollittMelanie2003Domf,CAT_20_2004astro.ph.11045J,CAT_21_Kato1987,CAT_22_1991ApJ...379...80K,CAT_23_2003A&A...406..579K,CAT_24_1998A&AS..133..129G,CAT_25_1999A&AS..139..359V,CAT_26_1982ApJ...252...81L,CAT_27_1979AJ.....84..725D,CAT_28_2010ApJ...714.1170M,CAT_29_2012ApJ...755...21M,CAT_30_2005Sci...307.1610G,CAT_31_2008ApJ...688.1029M,CAT_32_1996ApJ...458..194M,CAT_33_1995ApJ...445..624O, CAT_34, CAT_35_2005MNRAS.360.1305R,CAT_36_1983AJ.....88..518R, CAT_37, CAT_38_1981ApJS...45...97S,CAT_39_1980A&AS...39..379T,CAT_40_TAYLOR_RM_2009ApJ...702.1230T,CAT_41_Condon_1998,CAT_42_2011ApJ...728...97V, CAT_43_1993AJ....106..444W}}. Similar to \citet{EXTRAGAL_2015A&A...575A.118O}  we multiply the uncertainties of the Taylor et al. (sub-)data set \citep{CAT_40_TAYLOR_RM_2009ApJ...702.1230T} with 1.22 according to \citet{122_TAYLOR_2011ApJ...726....4S}. No additional polarimetric information such as e.g. Faraday spectra was used in this work. The data and the corresponding standard deviations are shown in Figs. \ref{fig:faraday_data} and \ref{fig:faraday_rms_data}. \\
Next, we specify the modeling of $\phi$ in Eq. \eqref{eq:faraday_data}. We will furthermore explain the treatment of the non-galactic components of $d_\phi$. 

\subsection{The model}
\label{subsec:the_model_faraday}

\subsubsection{The sky model}
\label{subsubsec:the_sky_model_faraday}

The results of the previous reconstruction \citep{EXTRAGAL_2015A&A...575A.118O} indicates that the Faraday sky has two characteristics that necessarily need to be modeled for an optimal inference. Firstly, the sign of the sky field changes rapidly, which reflects the rather abrupt directional changes of the LOS magnetic field component. Secondly, the absolute value of the signal varies over two orders of magnitude, most notably the Galactic disk will have a much stronger signal than the Galactic poles. For these reasons, we chose to parameterize $\phi: S^2 \rightarrow \mathbb{R}$ as a point-wise product between two fields living on the sky, or explicitly  
\begin{equation}
\label{eq:faraday_model}
\phi \equiv \chi e^\rho\, \frac{\mathrm{rad}}{\mathrm{m}^2},
\end{equation}
where $\chi: S^2 \rightarrow \mathbb{R}$ will be called the sign field, as it is able to capture the information on the sign of the Faraday depth, and $\rho: S^2 \rightarrow \mathbb{R}$ will be called the amplitude field, as $e^\rho$ is strictly positive after the exponentiation and is able to model the large amplitude variations of the Faraday sky over orders of magnitude. Both fields are unitless. This is a more generic approach as in \citet{OLD_FARADAY_2012A&A...542A..93O}, where instead of the amplitude field a profile function was used to capture the latitudinal dependence of the overall Galactic Faraday dispersion profile.
Note that the sign field is in no way constrained to only contain information on the sign of the signal, but also can capture morphological features, leading to a degeneracy between the two fields.
This could be broken by either imposing constraints on the correlation structure for (one of) the fields or by introducing new data that informs the algorithm on one of the two fields. However, it turns out that the symmetry is sufficiently broken by the Gaussian process priors we impose on both fields separately in combination with the specific functional form of $\phi$ as given by Eq. \eqref{eq:faraday_model}. Sign variations can only be captured by the sign field, ensuring that it is structured. The overall amplitudes of these fluctuations, which change as a function of position on the sky, are, however, more easily represented by the amplitude field, which therefore preferentially absorbs those.  
Both fields are assumed to have independent and Gaussian isotropic statistics. Their joint prior probability function is
\begin{equation}
\label{eq:gaussian_prior}
\mathcal{P}\left(\chi, \rho \vert  S_\chi, S_\rho\right) = \mathcal{G}\left(\chi , S_\chi\right) \mathcal{G}\left(\rho , S_\rho\right). 
\end{equation}
However, in contrast to the distribution in Eq. \eqref{eq:faraday_noise_pdf}, the covariances $S_\chi$ and $ S_\rho$ are unknown.
If this were not the case, the above model in a Bayesian setting would directly lead to a specific application of the well known (non-linear) Wiener filter \citep{Wiener, OLD_IFT_PhRvD..80j5005E}. This is not the case unfortunately, as in our case we have little a priori knowledge on the correlation structure. Similar inference problems have been solved in the past, e.g. in the previous reconstructions of the Faraday sky by \citep{OLD_FARADAY_2012A&A...542A..93O,EXTRAGAL_2015A&A...575A.118O}, but also e.g. in \citep{D3PO_2015A&A...581A.126S, DANIEL_2018A&A...610A..61P, REIMAR_2019arXiv190105971L}.  \\
All of these inferences relied on a framework laid out by information field theory (IFT) \citep{OLD_IFT_PhRvD..80j5005E, IFT_2018arXiv180403350E}. IFT connects Bayesian statistics with methods from statistical and quantum field theory, joining them into a inference scheme that connects noisy, incomplete data with the underlying continuous field(s). 
Considering the problem mentioned above, the requirement of simultaneously inferring the map and the correlation structure of a field leads to the critical filter formalism, which was first formulated in IFT by \citet{CRIT_FILT_0_2010PhRvE..82e1112E} and  \citet{CRIT_FILT_1_2011PhRvD..83j5014E}. We will follow the most modern formulation, as outlined in \citet{JAKOB_2017arXiv171102955K}. We will shortly summarize the modeling of the respective covariance matrices in the Appendix. 
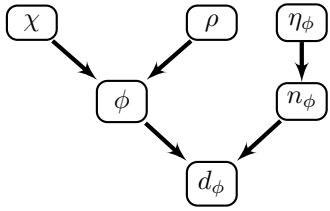
\begin{figure}
\centering
% \begin{center}
\begin{tikzpicture}[auto]
%    \small
    % node placement with matrix library
  \matrix[ampersand replacement=\&, row sep=0.5cm, column sep=0.5cm] {
    % Place nodes
    \node [block] (chi) {$\chi$};
    \&
    \&
    \node [block] (rho) {$\rho$};
    \&
    \node [block] (eta) {$\eta_\phi$};
    \&
    \&
    \\
    \&
    \node [block] (phi) {$\phi$};
    \&
    \&
    \node [block] (n) {$n_\phi$};
    \&
    \&
    \\
    \&
    \&
    \node [block] (d_phi) {$d_\phi$};
    \&
    \&
    \&
    \\ 
    }; 
    % Draw edges
    \path [line] (chi) -- node{}(phi);
    \path [line] (rho) -- node{}(phi);
    \path [line] (phi) -- node{}(d_phi);
    \path [line] (n) -- node{}(d_phi);
    \path [line] (eta) -- node{}(n);

%    \path [line] (B_rec) -- node {$\mathrm{(g)}$}(B_today);
\end{tikzpicture}

% \end{center}
\caption{\label{fig:sky_model_faraday} Hierarchical Bayesian model for the initially revised Faraday sky. We decompose the Faraday depth $\phi$ into the fields $\chi$ and $\rho$, which are supposed to capture the sign and the overall amplitude of the signal, respectively (see Eq. \eqref{eq:faraday_model}). The Faraday depth field $\phi$ together with the measurement noise $n_\phi$ determine the observed data $d_\phi$ via Eq. \eqref{eq:faraday_data}. The noise $n_\phi$ of every measurement $i$ is assumed to be drawn from a Gaussian with variance $\eta_i\sigma^2_i$, where $\sigma_i$ is the reported uncertainty and $\eta_i$ a unknown uncertainty correction fudge factor.}
\end{figure}    

\subsubsection{The noise model}
\label{subsubsec:the_noise_model_faraday}

As mentioned before, we also need to find an estimate for the non-galactic contributions to the Faraday rotation. In this part of the inference, we follow the approach in \citet{OLD_FARADAY_2012A&A...542A..93O} very closely. 
We will start by inserting Eq. \eqref{eq:faraday_separation} into the measurement equation \eqref{eq:faraday_data} and define an effective noise term $\tilde{n}_\phi $:
\begin{equation}
\label{eq:faraday_effective_noise}
d_\phi = \mathcal{R}_\phi (\phi_{\mathrm{gal}} + \phi_{\mathrm{etc}}) + n_\phi \equiv R_\phi\phi_{\mathrm{gal}} + \tilde{n}_\phi   
\end{equation}
We have no reason to drop the assumption of Gaussianity for the new noise term $\tilde{n}_\phi $, but have to adapt the covariance to include the increase in uncertainty. As we have no a priori guesses on the specific systematics of the different sources, we will infer this increase in uncertainty as well. 
This is implemented by introducing the following model 
\begin{equation}
\label{eq:noise_model}
\widetilde{N}_\phi = \mathrm{diag}\left(\eta_\phi \sigma^2_\phi\right),
\end{equation}
for the new noise covariance $\widetilde{N}$ where $\eta_\phi$ are the parameters that need to be inferred and $\sigma^2_\phi$ is the reported measurement uncertainty. We will assume the $\eta_\phi$ to follow an inverse gamma distribution:
\begin{equation}
\label{eq:noise_inference}
\mathcal{P}\left(\eta_\phi|\alpha_\phi, \beta_\phi\right) =\frac {\beta_\phi ^{\alpha_\phi }}{\Gamma (\alpha_\phi )}\eta_\phi^{-\alpha_\phi -1}\exp \left(-{\frac {\beta_\phi }{\eta_\phi}}\right) 
\end{equation}

The hyper-parameters $\alpha_\phi$ and $\beta_\phi$ need to be specified for the inference. In contrast to \citet{OLD_FARADAY_2012A&A...542A..93O} we do not set $\alpha_\phi = 1$ and $\beta_\phi$ such that the prior expectation value of $\ln(\eta_\phi)$ is zero, but demand a steeper slope with $\alpha_\phi = 2$ and the prior expectation value of $\eta_\phi$ to be unity. The reason for this is that we are allowing for more degrees of freedom in our new sky model as compared to \citet{OLD_FARADAY_2012A&A...542A..93O}, as we here have a two dimenensional amplitude field, where as \citet{OLD_FARADAY_2012A&A...542A..93O} only had a one dimensional profile function that was empirically determined without any prior. The field prior we have to impose on this amplitude field works against the data, increasing the tendency to classify data structures as noise. The larger value of $\alpha_\phi$ just counter balances this tendency to a degree that our resulting initially revised Faraday map (Fig. \ref{fig:ini_faraday_mean}) looks similar to that of \citet{EXTRAGAL_2015A&A...575A.118O} in Fig. \ref{fig:pre_faraday_mean}.

\subsection{The inference}
\label{subsec:the_inference_faraday}

The hierarchical Bayesian model described in the previous sections can be symbolically depicted as a hierarchical tree, of which the lower branches are shown in Fig. \ref{fig:sky_model_faraday}.
The higher branches of the tree contain the hyper-parameters and are discussed in the appendix together with the more explicit likelihood and prior functions. \\
The evaluation of the posterior is a non-trivial task in our case, as the non-linearities in the models lead to highly non-Gaussian distributions for which no analytical description is known to exist. Furthermore, the fact that we aim for an simultaneous inference of the fields determining the Faraday map as well as their correlation structures leads to a strong interdependence between all degrees of freedoms. Together with the high dimensionality of the problem with of the order of $10^6$ unknown parameters for the desired resolution of $\approx 10$ arcmin this leads to considerable numerical complexity.\\
IFT solves such problems by providing efficient and well tested algorithms for posterior evaluation. In contrast to \citet{OLD_FARADAY_2012A&A...542A..93O}, we employ here the Metric Gaussian Variational Bayes method (MGVI), which is better suited for simultaneous map and correlation structure inference compared to the maximum a posteriori algorithm used in \citet{OLD_FARADAY_2012A&A...542A..93O}, see \citet{JAKOB_2017arXiv171102955K} for details. We furthermore implement the aforementioned hierarchical tree as a forward model, which aims among other things for a more efficient numerical coupling between different parts of the model and allows the employment of sampling methods \citep{JAKOB_2017arXiv171102955K}. \\
On a technical side, the inference was conducted using the newest version (version 5) of the NIFTy python package \citep{NIFTY_2013A&A...554A..26S, NIFTY_2017arXiv170801073S, NIFTY} for numerical information field theory. NIFTy is specifically designed for IFT applications and provides libraries for implementing the above mentioned models as well as for inferring their parameters and the uncertainties of those.

\subsection{The results}
\label{subsec:results_faraday}

\subsubsection{Comparison to older results}
\label{subsubsec:comaprison_to_older_results}

\begin{figure}
\begin{subfigure}{\linewidth}
\centering
\includegraphics[width=1.0\textwidth]{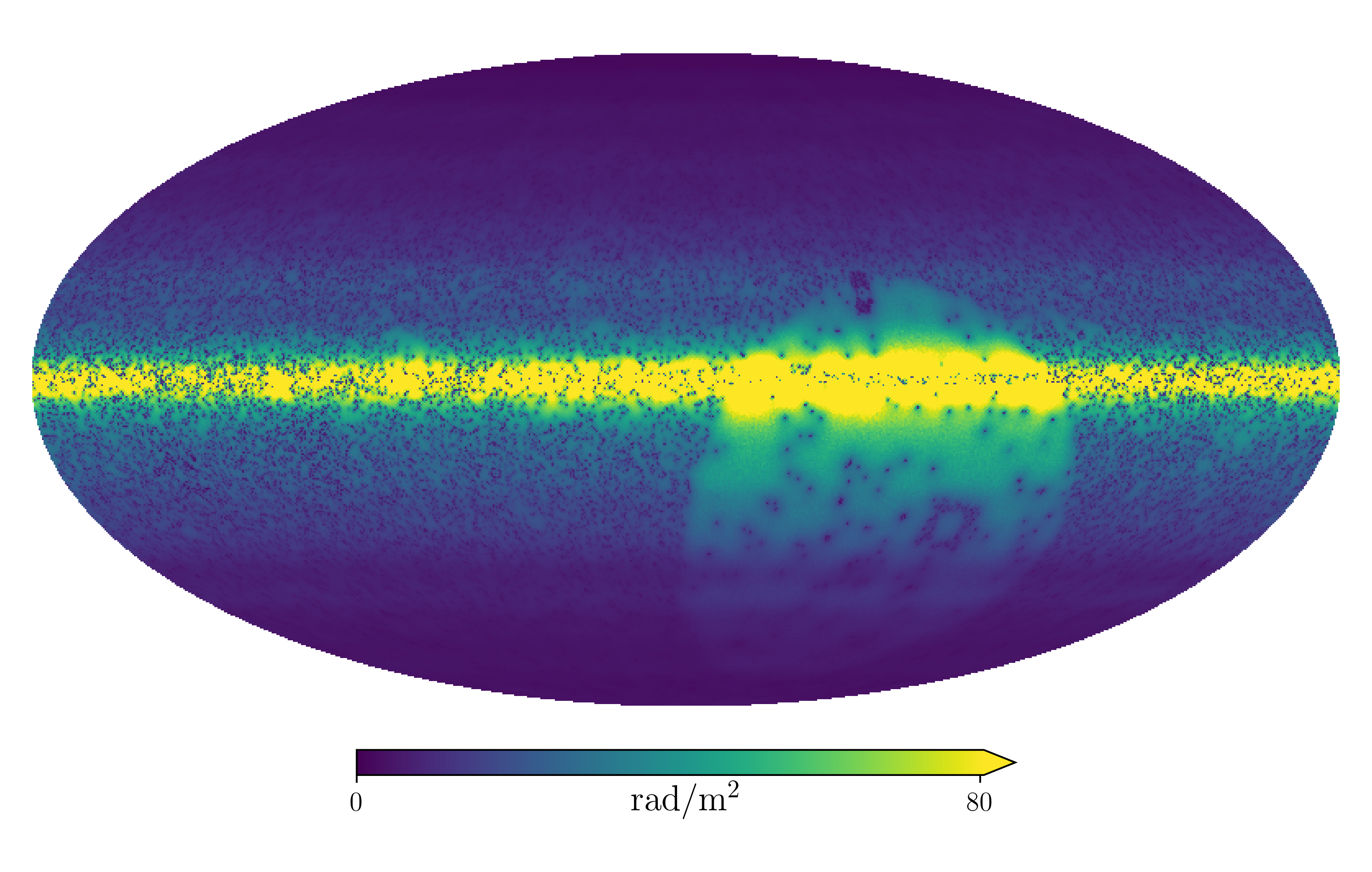}
\caption{\label{fig:pre_faraday_var}}
\end{subfigure}
\begin{subfigure}{\linewidth}
\centering
\includegraphics[width=1.0\textwidth]{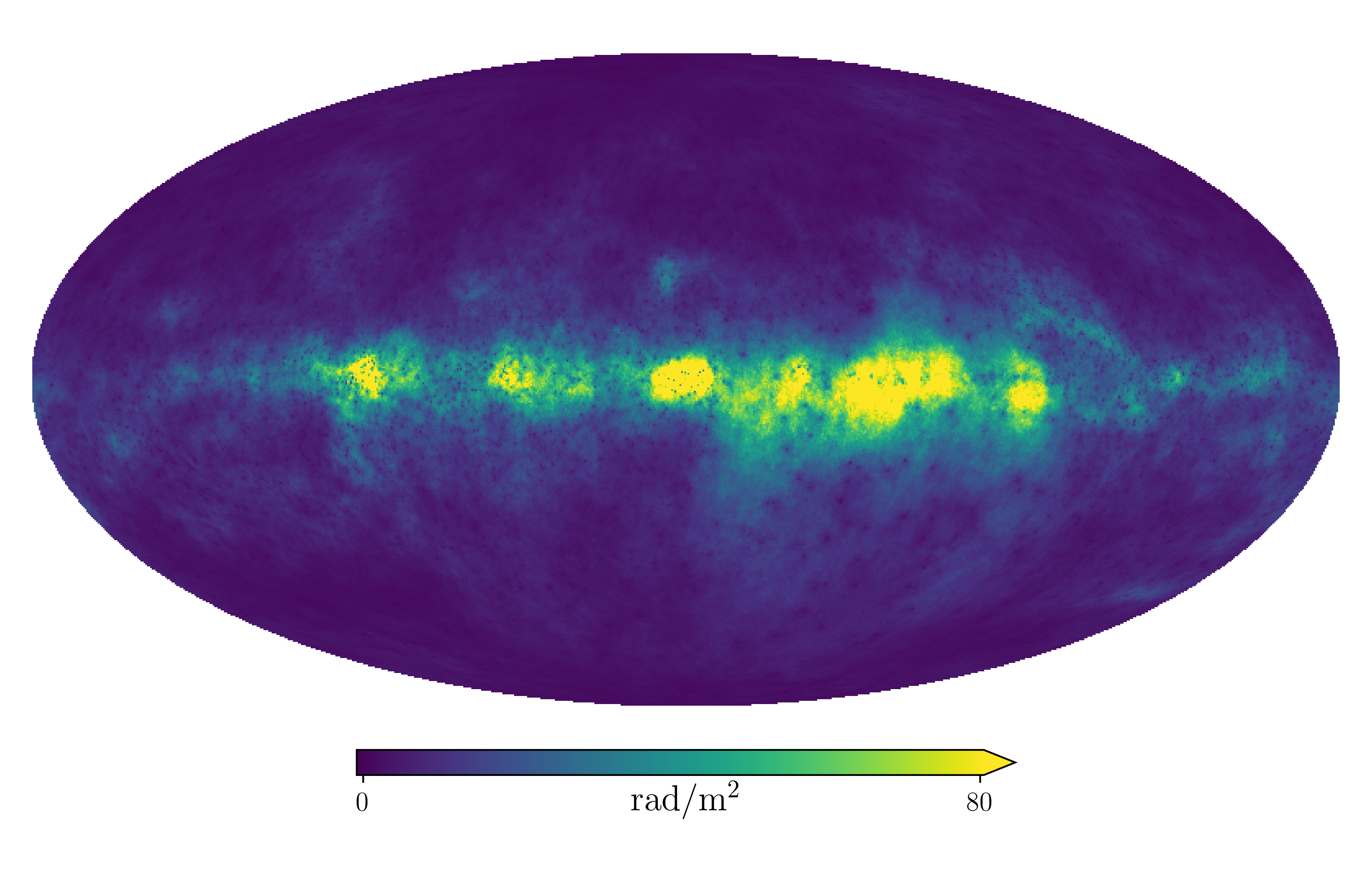}
\caption{\label{fig:ini_faraday_var}}
\end{subfigure}
\begin{subfigure}{\linewidth}
\centering
\includegraphics[width=1.0\textwidth]{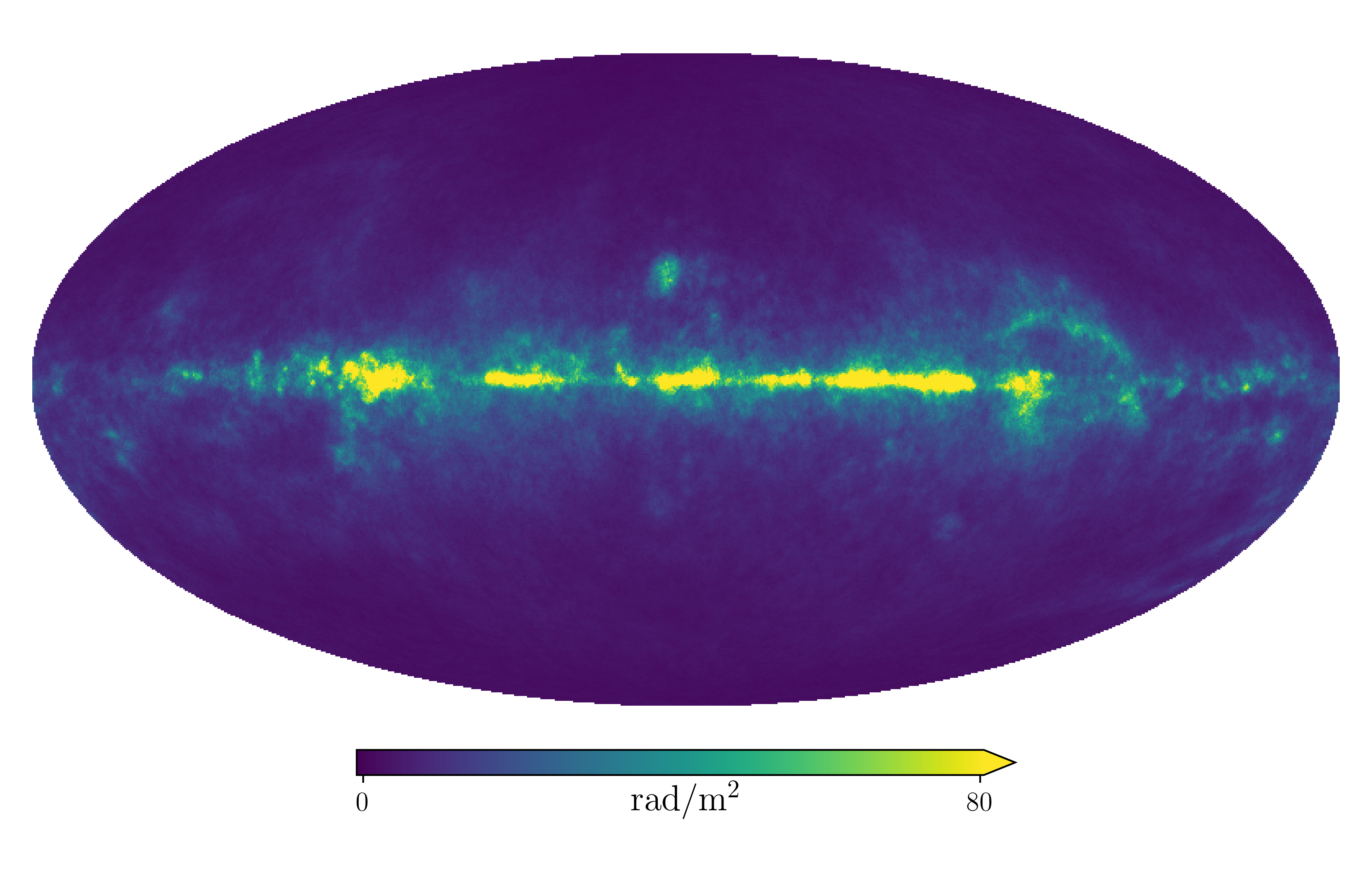}
\caption{\label{fig:rev_faraday_var}}
\end{subfigure}
\caption{\label{fig:faraday_var} Uncertainties of the different reconstructions. Figure (a) is the uncertainty corresponding to the previous reconstruction of \citet{EXTRAGAL_2015A&A...575A.118O} shown in Fig. \ref{fig:pre_faraday_mean}. Figure (b) shows the uncertainty corresponding to the initially revised reconstruction shown in Fig. \ref{fig:ini_faraday_mean}. Figure (c) finally shows the uncertainty corresponding to the revised reconstruction including the free free data shown in Fig. \ref{fig:rev_faraday_mean}.}
\end{figure}

The initially revised Faraday depth map relying on rotation measure (RM) data only and inferred according to the model depicted in Fig. \ref{fig:sky_model_faraday} is shown in Fig. \ref{fig:ini_faraday_mean}, the previous one of \citet{EXTRAGAL_2015A&A...575A.118O} in Fig. \ref{fig:pre_faraday_mean}, with corresponding standard deviations in Figs. \ref{fig:ini_faraday_var} and \ref{fig:pre_faraday_var}, respectively. The difference of both maps are shown in Fig. \ref{fig:diff_pre_ini}. A visual comparison of the two maps reveals satisfying agreement. The difference map, however, shows that there are indeed significant small scaled deviations of the maps in the area of the Galactic disk. These may be a result of the different noise estimation parameters in the updated inference and/or the different treatment of the amplitude field (see the discussion in Sec. \ref{subsubsec:the_noise_model_faraday}). One should note that the revised reconstruction was conducted with a doubled resolution, which may also be responsible for some of the very small scaled differences.\\ 
The uncertainty maps nicely demonstrate the impact of the sky modeling on the inference. Fig. \ref{fig:pre_faraday_var} shows that the uncertainty of the previous reconstruction, which exhibits imprints of the longitudinal galactic profile used in the amplitude modeling. Our updated uncertainty map in Fig. \ref{fig:ini_faraday_var} shows a decrease in uncertainty in certain regions, especially in the galactic disk. This simply reflects the fact that our updated inference scheme can use the correlation information encoded in the data much more efficiently.

\begin{figure}
\begin{minipage}{\linewidth}
\centering
\includegraphics[width=1\textwidth]{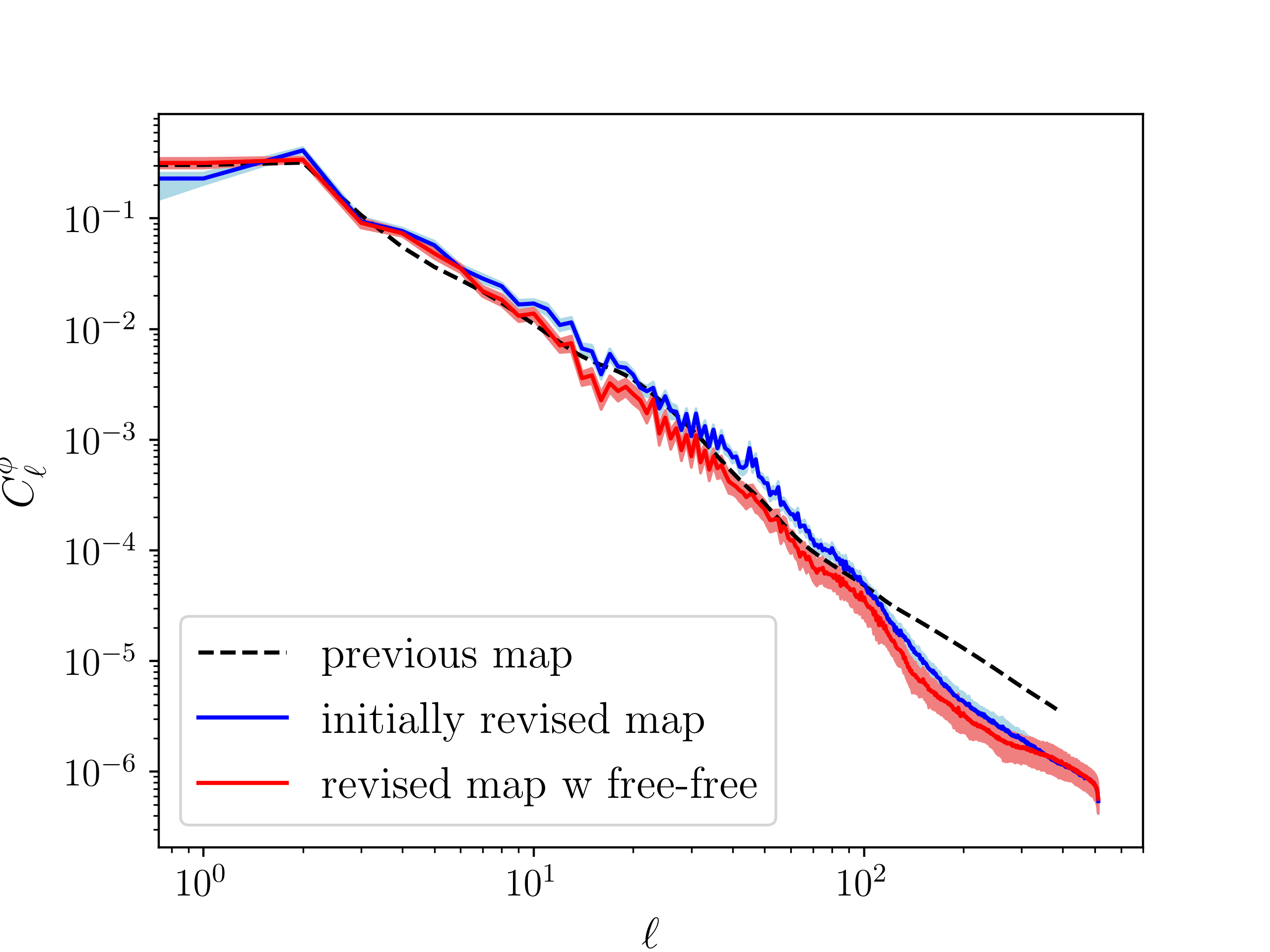}
\caption{\label{fig:power} Comparison of the power spectrum inferred by \citet{EXTRAGAL_2015A&A...575A.118O} (dashed black) to the spectrum of the revised Faraday map including the free-free EM data (red) and the results of the initially revised inference in Sec. \ref{sec:improving_the_inference_of_the_faraday_sky} (blue). Both of the latter inference algorithms also produce uncertainties in the power spectrum, which are depicted as shaded areas in the respective colors.}
\end{minipage}
\end{figure}
\begin{figure}
\begin{subfigure}{\linewidth}
\centering
\includegraphics[width=1\textwidth]{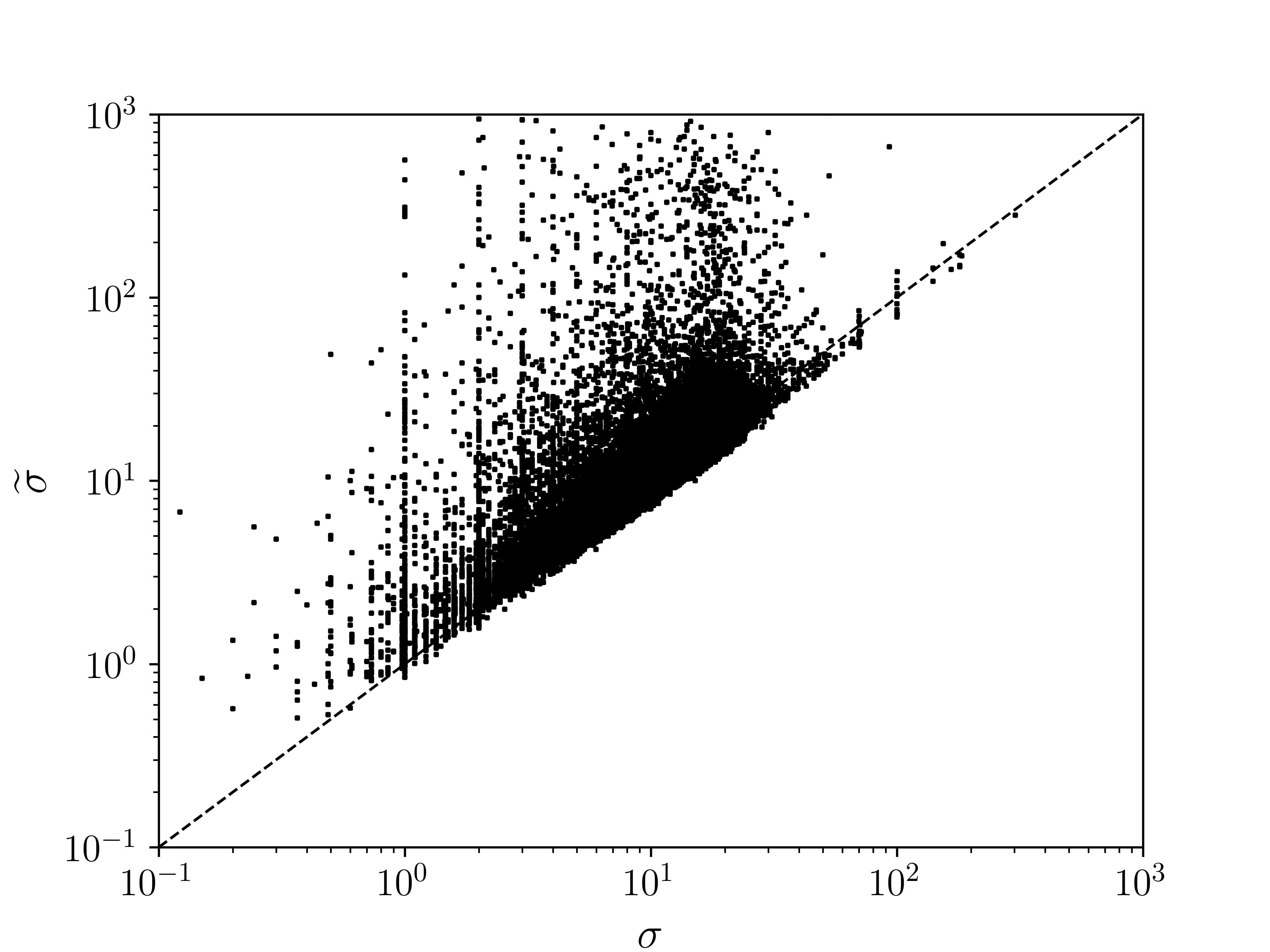}
\caption{\label{fig:noise_scatter}}
\end{subfigure}
\begin{subfigure}{\linewidth}
\centering
\includegraphics[width=1\textwidth]{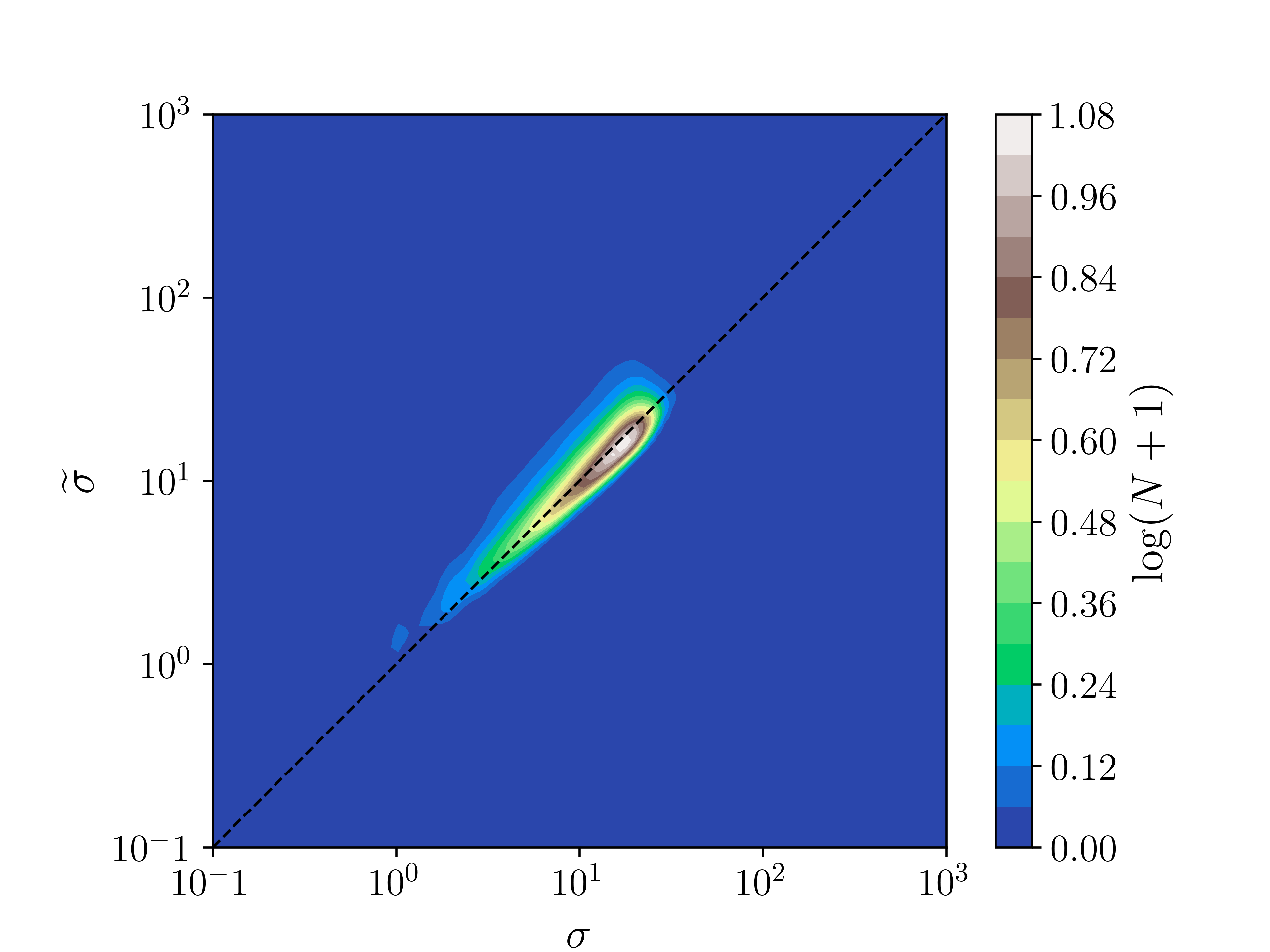}
\caption{\label{fig:noise_contour}}
\end{subfigure}
\caption{\label{fig:noise} Plots describing the impact of the noise estimation procedure. Figure (a) shows the estimated noise standard deviations compared to the observed ones. Note that the visual impression is dominated by the outliers. Figure (b) shows the estimated noise standard deviations compared to the observed ones. This plot is based the same information as Fig. \ref{fig:noise_scatter}, but shows the underlying probability density function estimated via Gaussian kernels. In both plots, equality between the two quantities is marked by the dashed line.}
\end{figure}

A further quantity which needs to be inspected are the power spectra depicted in Fig. \ref{fig:power}, as they have been inferred with different algorithms in the previous and initially revised approaches. We note that we can confirm the slope of the spectrum. The revised map has significantly less power on small scales. 
Again, the parameter change in the noise estimation may be the reason for this discrepancy, as we know that it can have a considerable impact on the smaller scales, as we change the ability of the algorithm to increase uncertainties to accommodate outliers in the data.\\
With respect to the error estimation, we show in Fig. \ref{fig:noise_scatter} the comparison of the reported observational noise standard deviation $\sigma$ and our estimate for the noise $\widetilde{\sigma}$ according to Eq. \eqref{eq:faraday_effective_noise}. The uncertainties of some sources have been multiplied by factors of up to $1000$, indicating strong internal or extra-galactic contributions to these particular RM's. Nonetheless, we confirm or even decrease the uncertainties for most measurements, as shown in Fig. \ref{fig:noise_contour}, where we show a normalized density plot in the $\sigma - \widetilde{\sigma}$ plane.\\
With this we conclude the comparison between the previous and the initially revised results and turn our face to the analysis of the underlying components of our sky model defined in Eq. \eqref{eq:faraday_model}.

\subsubsection{The sign and amplitude fields}
\label{subsubsec:the_sign_amplitude_fields}

\begin{figure}
\begin{subfigure}{0.99\linewidth}
\centering
\includegraphics[width=1.0\textwidth]{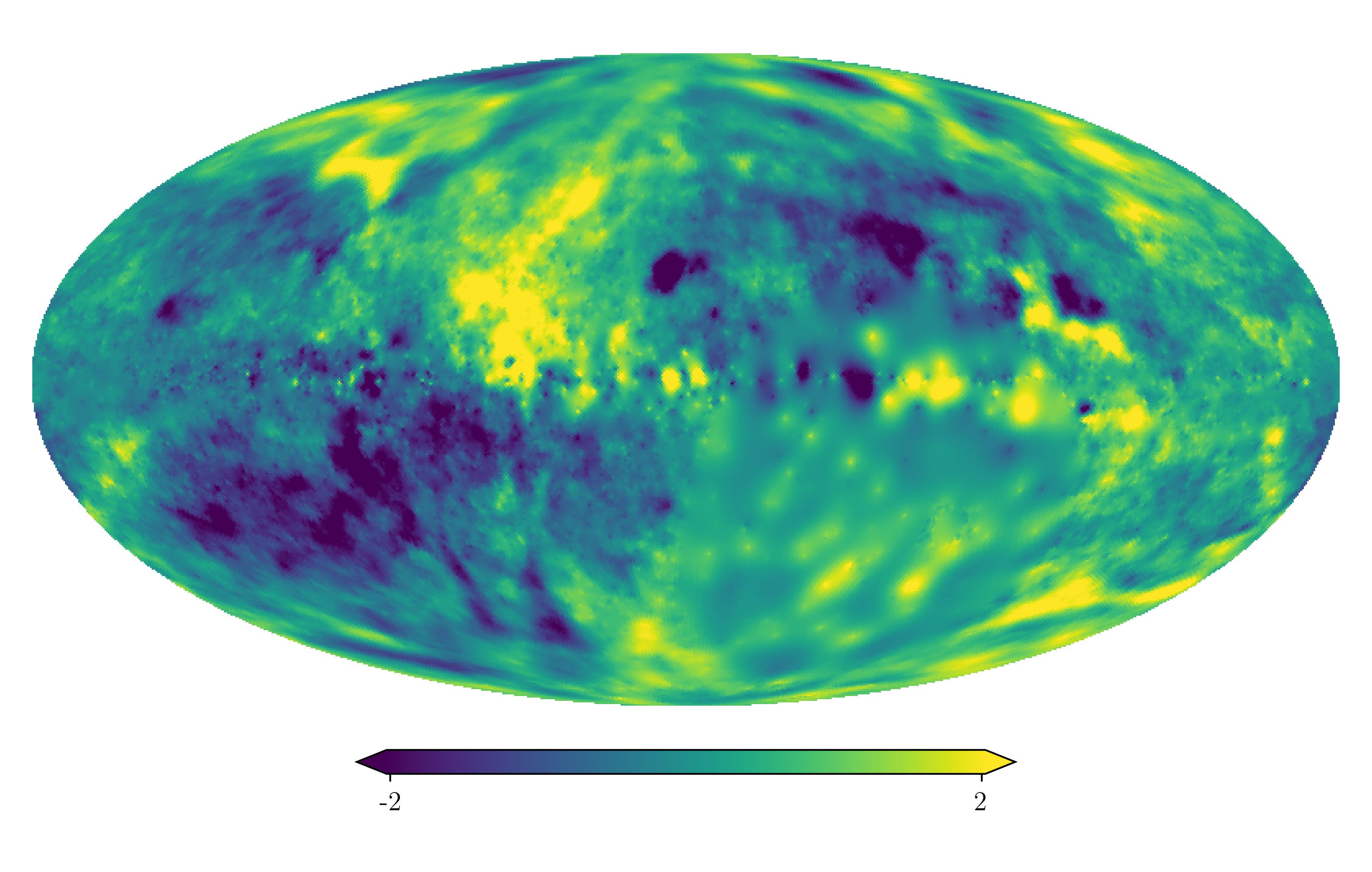}
\caption{\label{fig:pre_sign}}
\end{subfigure}
\begin{subfigure}{0.99\linewidth}
\centering
\includegraphics[width=1.0\textwidth]{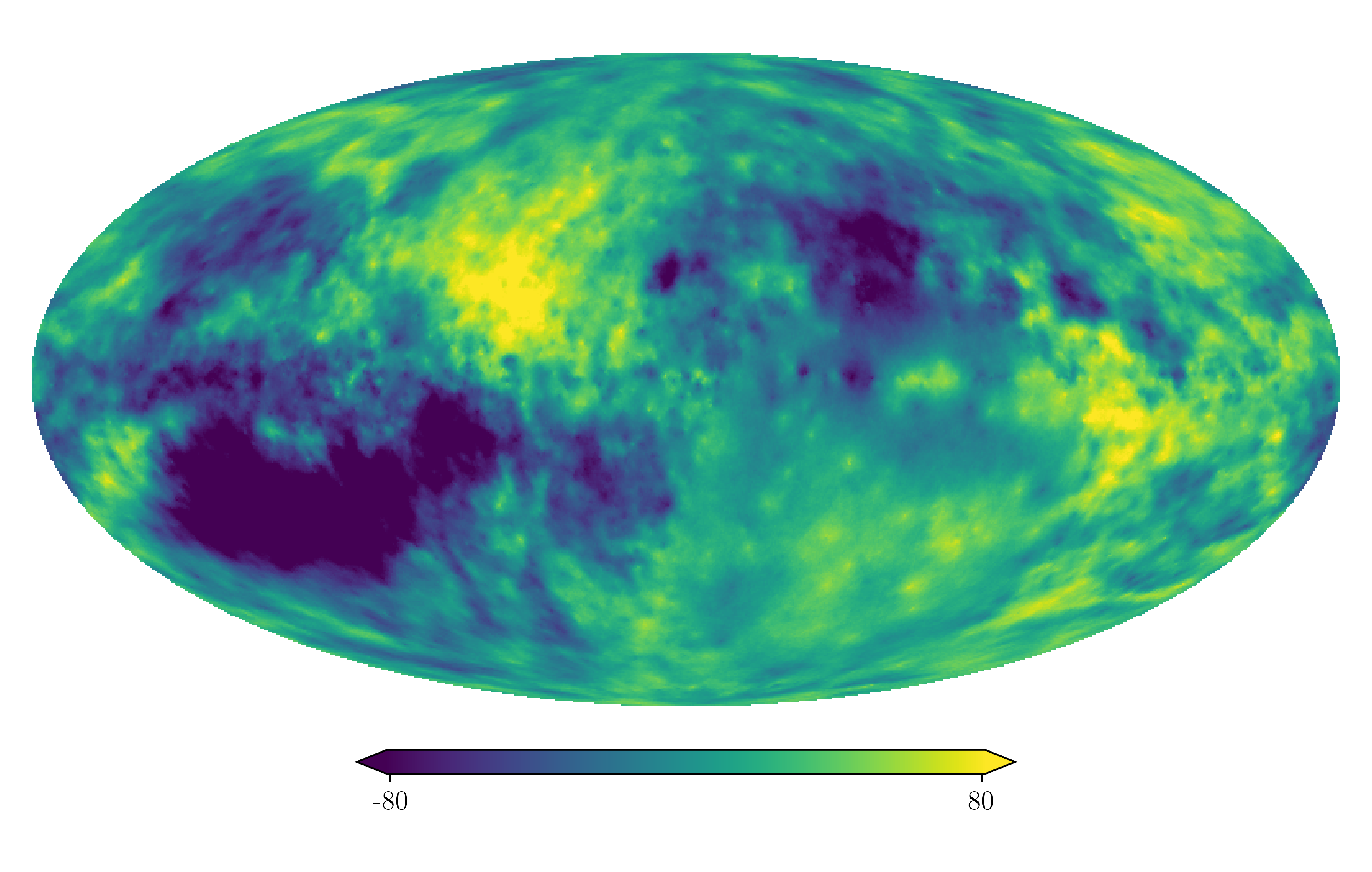}
\caption{\label{fig:ini_sign}}
\end{subfigure}
\begin{subfigure}{0.99\linewidth}
\centering
\includegraphics[width=1.0\textwidth]{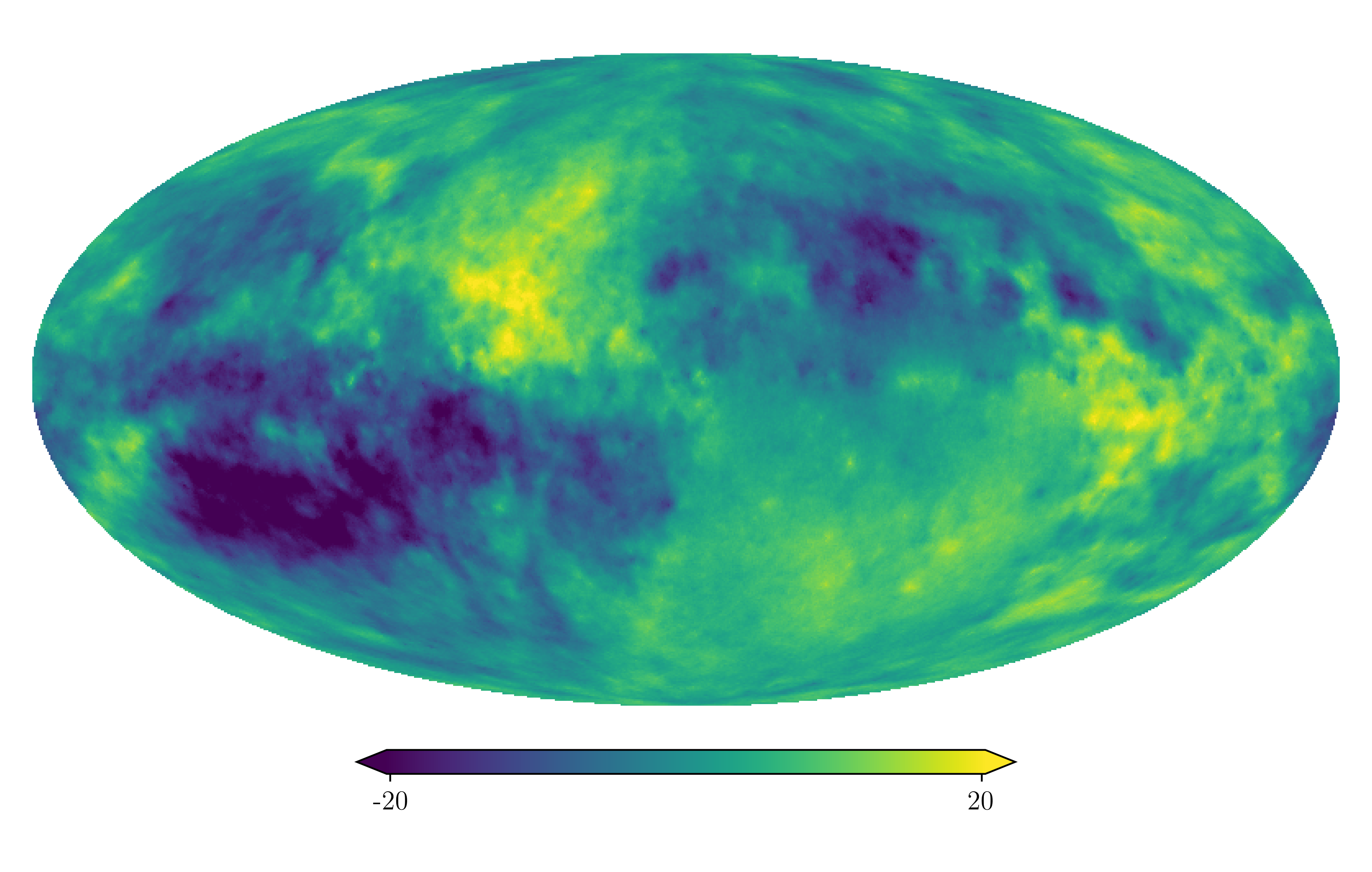}
\caption{\label{fig:rev_sign}}
\end{subfigure}
\caption{\label{fig:sign} Sign fields of the different reconstructions. Figure (a) shows the field $s$ of the previous reconstruction by \citet{EXTRAGAL_2015A&A...575A.118O}. Figure (b) shows the field $\chi$ of the initially revised reconstruction. Figure (c) finally shows the field $\chi$ of the revised reconstruction including the free free data. Note that the difference in the scales of these fields is compensated by corresponding differences in the respective Faraday amplitude fields.} 
\end{figure}

\begin{figure*}
\begin{subfigure}{0.49\linewidth}
\centering
\includegraphics[width=1.0\textwidth]{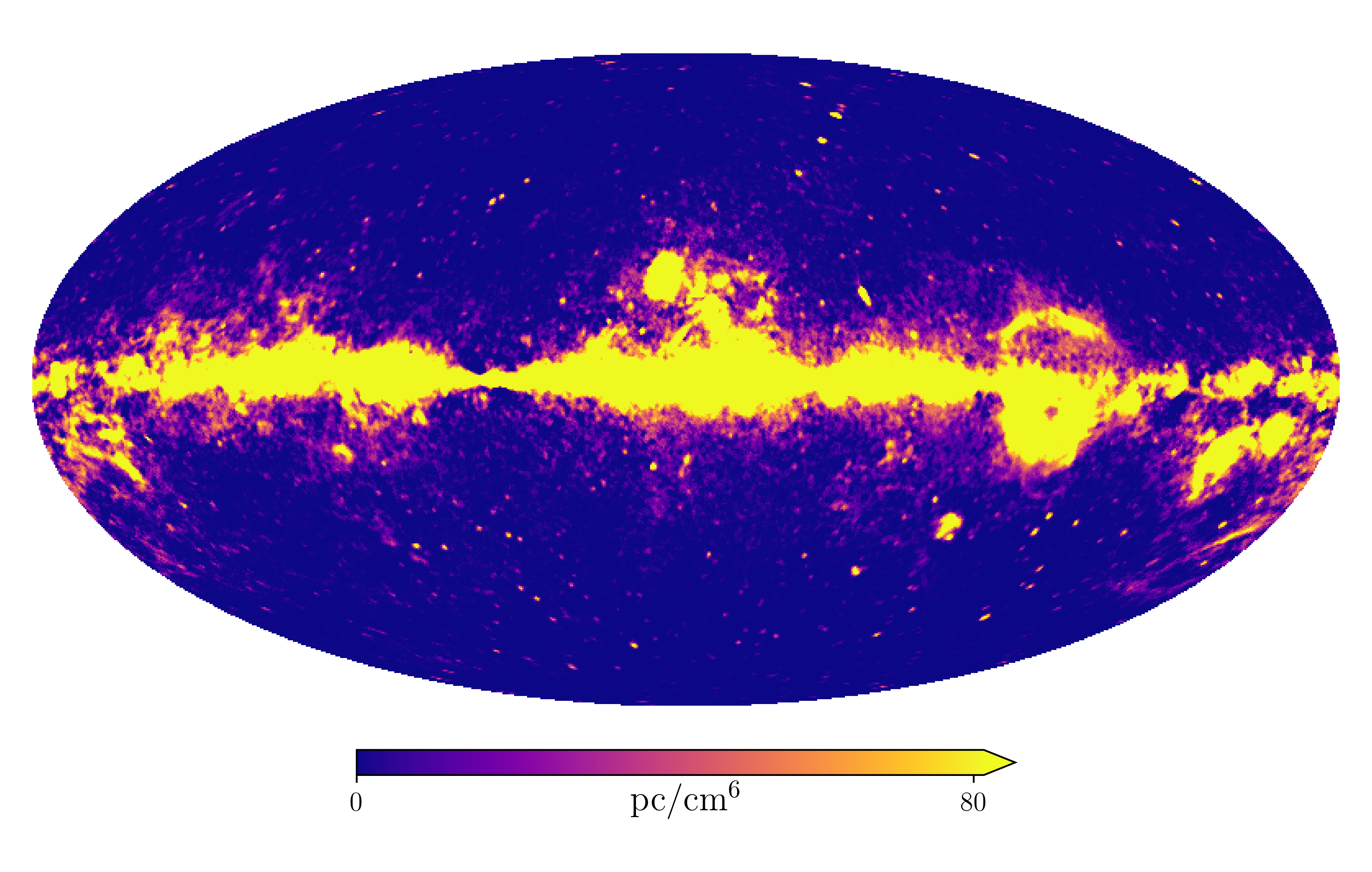}
\caption{\label{fig:planck}}
\end{subfigure}
\begin{subfigure}{0.49\linewidth}
\centering
\includegraphics[width=1.0\textwidth]{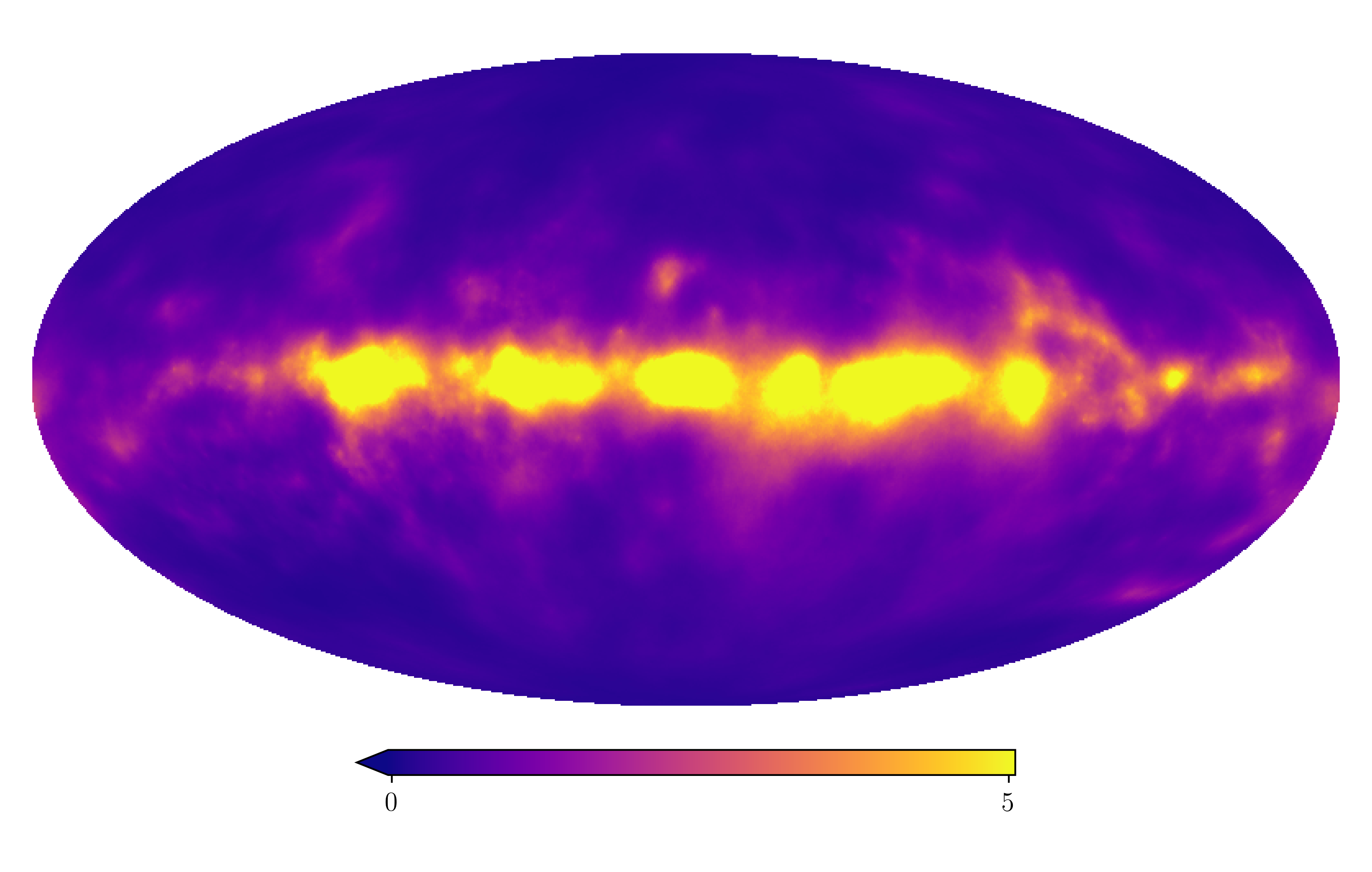}
\caption{\label{fig:ini_amplitude}}
\end{subfigure}
\begin{subfigure}{0.49\linewidth}
\centering
\includegraphics[width=1.0\textwidth]{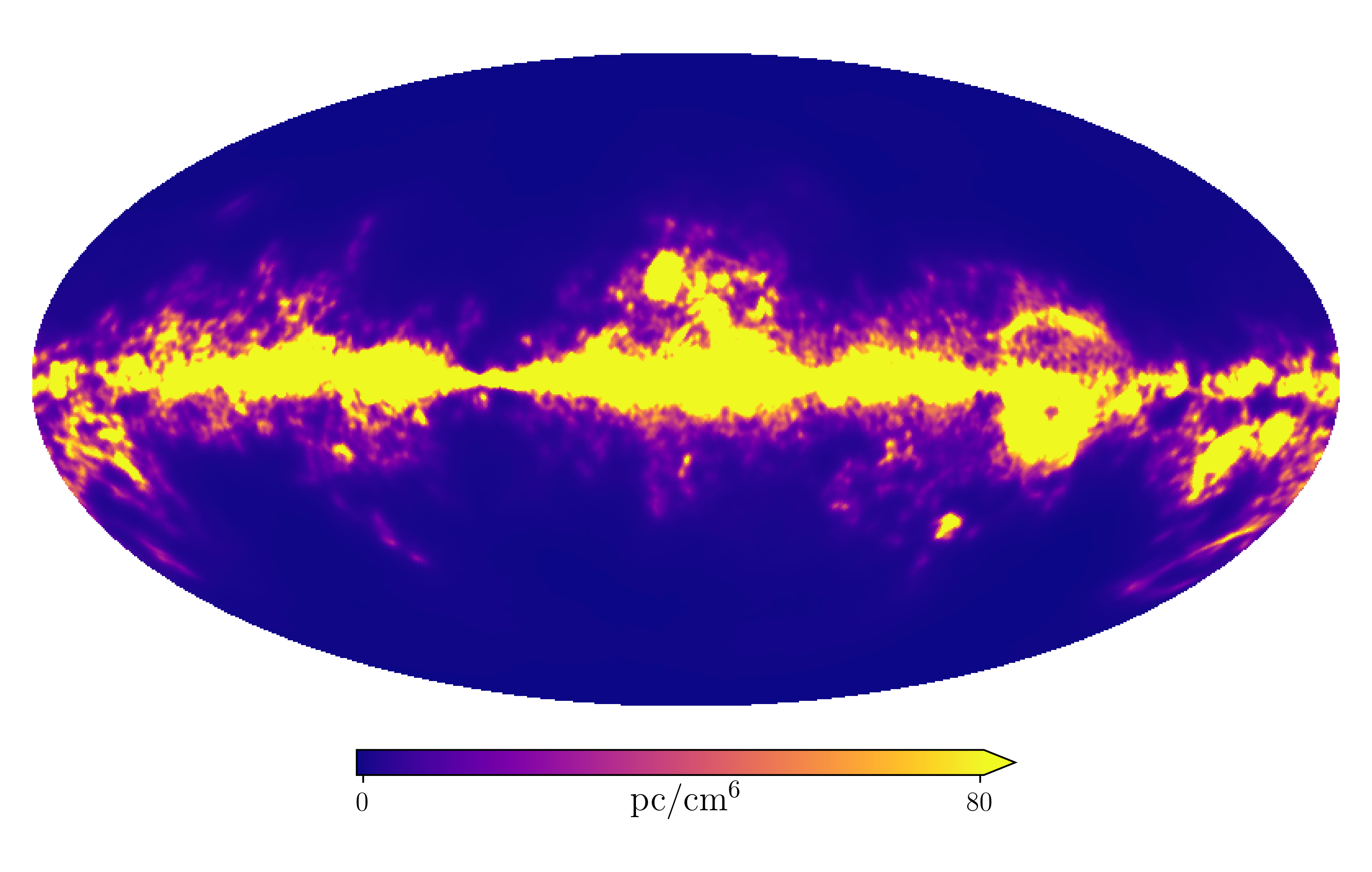}
\caption{\label{fig:rev_free_free}}
\end{subfigure}
\begin{subfigure}{0.49\linewidth}
\centering
\includegraphics[width=1.0\textwidth]{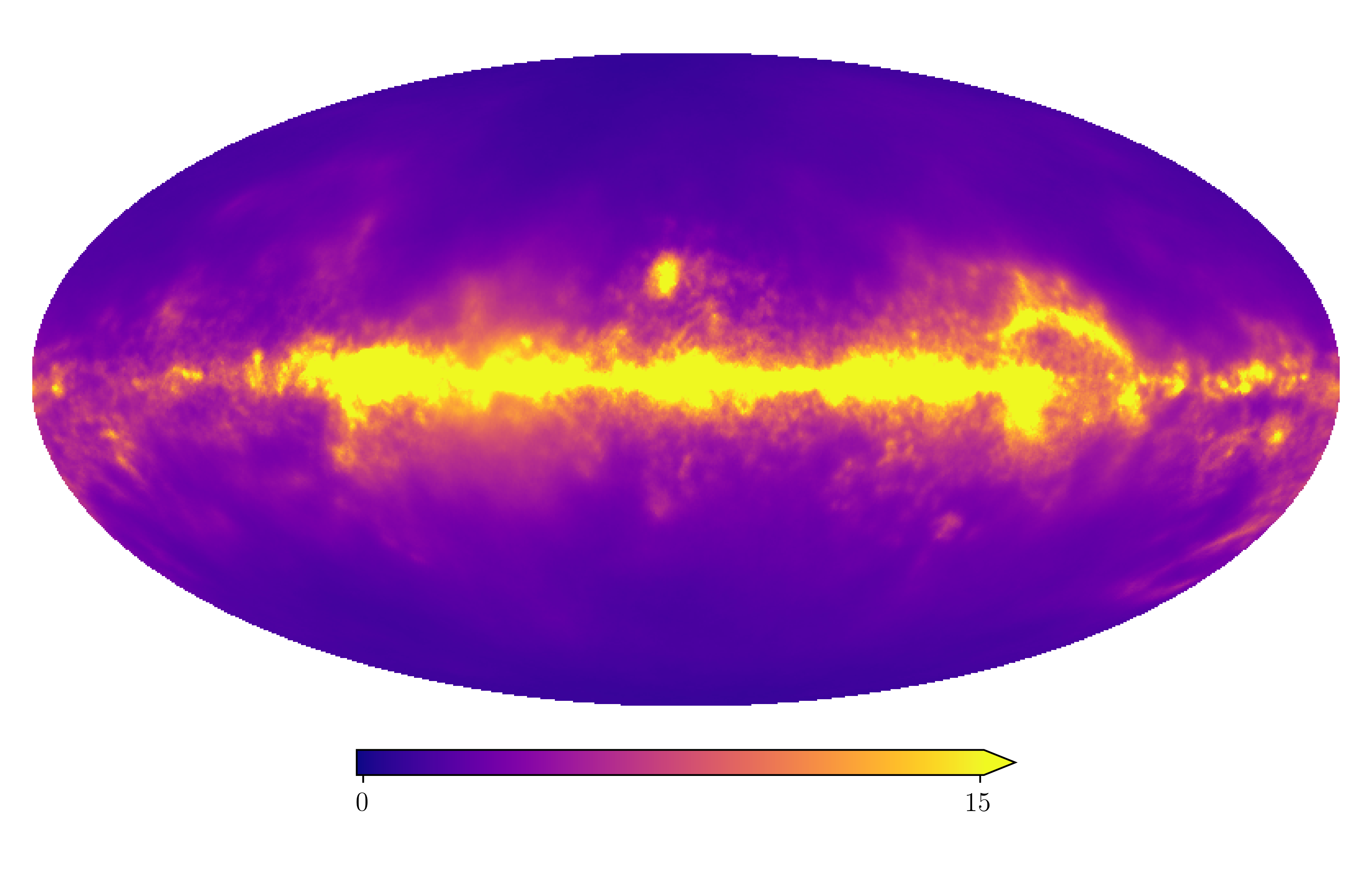}
\caption{\label{fig:rev_amplitude}}
\end{subfigure}
\caption{\label{fig:amplitudes} Free-free data, reconstructions and Faraday amplitude fields. Figure (a) shows the  Galactic free-free EM map as obtained by the Planck Collaboration \citep{PLANCK_FOREGROUND_2016A&A...594A..10P}. Figure (b) shows the exponentiated amplitude field $\rho$ of the initially revised reconstruction. This field is not constrained by free-free data. Figure (c) shows the exponentiated $\epsilon$ field. This field is part of the revised Faraday reconstruction as well as the denoised free-free sky. A logarithmic version of this plot is shown in Fig. \ref{fig:rev_epslion}. Figure (d) shows the Faraday amplitude field of the revised reconstruction including free-free data. This field is part of the result of the model in Eq. \eqref{eq:new_faraday_model}. A logarithmic version of this plot is shown in Fig. \ref{fig:rev_log_amplitude}.} 
\end{figure*}

The component fields $\chi$ and $\rho$ are shown in Figs. \ref{fig:ini_sign} and \ref{fig:ini_amplitude}, respectively. The sign field captured most of the small scale structure, but has more or less completely lost all information on the Galactic disc profile, which was absorbed by the amplitude field, as intended by our modeling. $\chi$ shows some similarity to the corresponding signal field inferred by \citet{EXTRAGAL_2015A&A...575A.118O}, which is shown in Fig. \ref{fig:pre_sign}.
The exponentiated amplitude field appears to be relatively smooth with few distinguished features apart from the large scaled disk profile. These features, however, show remarkable similarity to galactic free-free emission maps, which was precisely measured by missions investigating the Cosmic Microwave Background (CMB) such as Planck \citep{PLANCK_FOREGROUND_2016A&A...594A..10P}, as it is a important foreground component in the microwave sky. We show the Planck free-free EM map in Fig. \ref{fig:planck}. The color scales were chosen to highlight the structures in the maps of which we think that they originate from the same galactic structures. This apparent resemblances in both maps have motivated us in including the free-free map in our inference.
A investigation of the physical plausibility of this correlation is given in the next section. 

\section{Including the Free-Free emission}
\label{sec:including_free_free}

\subsection{The physics}
\label{subsec:the_physics_free_free}
In addition to the visual similarities between the free-free map and our inferred amplitude field, we want to further motivate this apparent congruence (and deviations from it) by investigating the physics of the free-free emission and the Faraday rotation a bit more closely. The $\mathrm{EM}$ traced by the bremsstrahlung emitted by thermal electrons on protons in the warm interstellar medium can be quantified via (see e.g. \citet{FREE_FREE_1998astro.ph..1121S})
\begin{equation}
\label{eq:free_free}
\mathrm{EM}_\mathrm{ff}\left(n_\mathrm{th}\right) =  \int_\mathrm{LOS} dl\, n_\mathrm{th}^2 \equiv \int_\mathrm{LOS} dl\, \mathrm{em}_\mathrm{ff},
\end{equation}
where we defined $ \mathrm{em}_\mathrm{ff} \equiv n^2_\mathrm{th}$ as a source term for $\mathrm{EM}_\mathrm{ff}$.\\
If we turn to the Faraday depth and want to study Eq. \eqref{eq:faraday_physics} with the intention of an comparison to $\mathrm{EM}_\mathrm{ff}$, we must quantify the dependence of $B_\mathrm{LOS}$ on the thermal electron density. In general, we can assume the magnetic flux to be frozen into the ISM \citep{FLUX_FREEZING_2006AIPC..875..289D}. The exact nature of the $B - n_\mathrm{th}$ relationship, however, is complicated as it strongly depends on the morphology and dynamics of the plasma under consideration. In the simplest case of an isotropically collapsing structure, flux freezing leads to an $\vert B\vert \propto n_\mathrm{th}^{\frac{2}{3}}$ dependence \citep{FLUX_FREEZING_3_1966MNRAS.133..265M}. For a realistic description the Galactic interstellar medium (ISM), however, one needs to consider non-linear magneto-hydrodynamics (MHD). Under certain simplifying assumptions as shown in \citet{FLUX_FREEZING_5} one may decompose the ISM flow into several wave types corresponding to different modes, leading to different density dependencies of the magnetic field strength. Most notably in the low density regime, the dominating mode approaches a constant in $n_\mathrm{th}$, implying no correlation between $\vert B \vert$ and $n_\mathrm{th}$ in this regime. Furthermore, as waves corresponding to different modes may very well travel trough the same regions within a short time span, the  $B - n_\mathrm{th}$ correlation may be subject to large variability in all regimes of measured electron density \citep{FLUX_FREEZING_2_2015ASSL..407..401V}.
We therefore assume the dependence of the magnetic field strength on $n_\mathrm{th}$ to follow a power law with an unknown and spatially dependent coefficient $\vert B\vert \propto n_\mathrm{th}^{p_B}$.
We can then rewrite the proportionality of the absolute value of integrand of the Faraday depth in Eq. \eqref{eq:faraday_physics} on $n_\mathrm{th}$
\begin{align}
\label{eq:free_free_faraday}
   B_\mathrm{LOS} n_\mathrm{th} &= \vert B\vert \cos(\theta) n_\mathrm{th} \propto \nonumber \\ 
 &\propto n_\mathrm{th}^{1 + p_B} \cos(\theta) \propto \mathrm{em}_\mathrm{ff}^{\frac{1+p_B}{2}} \cos(\theta).
\end{align}
Here we introduced the angle $\theta$ indicating the orientation of the magnetic field with respect to the LOS. Its impact will be discussed later on.
If $p_B \approx 1$ for all locations along a line of sight, the absolute value of the Faraday integrand does strongly depend on the free-free emissivity and the additional electron density dependence has completely canceled. In the unrealistic case that $p_B = -1$, the absolute value of the Faraday integrand would be completely uncorrelated to the free-free source term. Furthermore as discussed before, in some cases the electron density will not be a good tracer for the magnetic field strength and the LOS projection might partly mask the relation in the observables, which again limits the above relationship.
Nonetheless, we think that the above considerations motivate the inclusion of the free-free map as a proxy for the Faraday amplitude field in our inference, as for most realistic cases of the power law index  $p_B$ the integral over both sides of Eq. \eqref{eq:free_free_faraday} will lead to a correlation between the free-free and the Faraday sky. On the other hand this analysis also indicates that the effects of ISM dynamics on the magnetic field strength have to be considered for a reliable inference. \\ 
A second effect that needs to be taken into account are sign reversals of the magnetic field along the line of sight, which could strongly affect the value of $\phi$ in contrast to $\mathrm{EM}_\mathrm{ff}$.
To consider these, we approximate the line of sight integral via
\begin{align}
\label{eq:field_geometry}
\phi &\propto  \int_\mathrm{LOS} dl\, \vert B \vert \cos(\theta) n_\mathrm{th} = L\langle \vert B \vert n_\mathrm{th} \cos(\theta)   \rangle_L \approx \nonumber \\
 &\approx L^2 \langle \vert B \vert n_\mathrm{th} \rangle_L \langle \cos(\theta)\rangle_L, 
\end{align}
where $L$ is the length of the line of sight and $\langle \ldots \rangle_L \equiv \frac{1}{L} \int_\mathrm{LOS} dl \ldots $ indicates an averaging process along the line of sight. The last approximation implies statistical independence between the magnetic field strength and $\theta$.
This shows that to good approximation, the geometry of the magnetic field can lead to a multiplicative term, which is not constrained by the free-free data alone and has to be considered independently. In our setup, it can only be further determined by discrepancies between the Faraday and free-free data sets.\\
Similar to the first part of this paper, we will first discuss properties of the EM data and then show the modeling of the respective sky maps in Sec. \ref{subsec:the_model_free_free}.

\subsection{The data}
\label{subsec:the_data_free_free}
In this work, we use the free-free EM map of the Planck 2015 release \citep{PLANCK_FOREGROUND_2016A&A...594A..10P}.  This map is a side product of a foreground subtraction process of the Planck team in their pursuit of the CMB sky reconstruction. The free-free sky is a dominant component of this foreground in the low frequency regime, along side with synchrotron and spinning dust emission \citep{LOW_DIFFUSE_2016A&A...594A..25P}. Most components in the microwave sky are hardly distinguishable, as they result from similar structures in the sky. Planck therefore employs a sophisticated component separation algorithm called $\mathtt{COMMANDER}$ \citep{COMMANDER_2008ApJ...676...10E}, mostly exploiting the different energy spectra of the components. 
This algorithm uses the Gibbs sampling technique to approximate the posterior distribution for all components, also making use of the WMAP data \citep{WMAP_2013ApJS..208...20B} and a 408 MHz survey map \citep{HASLAM_2015MNRAS.451.4311R} for the separation.
An updated version of $\mathtt{COMMANDER}$ employed in the 2018 release of Planck takes the correlation structure of the components into account. However, Planck has not released an updated free-free map in 2018.\\
As a side product of the estimation of the posterior via sampling, Planck provides us not only with the posterior mean of the free-free component, but also with uncertainties, which are a valuable source of information and lets us treat the free-free data as a slightly noisy measurement of the true free-free $\mathrm{EM}$ with known uncertainties. Unfortunately, upon inspecting the Planck data, we found some pixels to have zero variances. As this is rather surprising given the large degeneracy between the components, we expect this to be a numerical artifact of the sampling procedure or of the marginalization process that results in the uncertainties for specific components. Our algorithm relies on accurate uncertainty measures for all pixels, as in case of a severe underestimation of the errors in some specific area, this region is over-weighted by the algorithm in the inference of the correlation structure, thereby immediately affecting other parts of the sky and subsequently all other maps of the inference. This has to be accounted for if one is not completely sure that the noise values are free of systematics. We will therefore perform a similar noise estimation procedure as for the Faraday data.\\
With this in mind, we write the measurement equation for the free-free data as
\begin{equation}
\label{eq:free_free_data}
d_{\mathrm{ff}} = \mathcal{R}_{\mathrm{ff}}\mathrm{EM}_\mathrm{ff} + n_{\mathrm{ff}, \mathrm{obs}} + n_{\mathrm{ff} , \mathrm{sys}} \equiv \mathcal{R}_{\mathrm{ff}} \mathrm{EM}_\mathrm{ff}+ \widetilde{n}_{\mathrm{ff}}.    
\end{equation}
In case of Planck, the response $\mathcal{R}_{\mathrm{ff}}$ is just a unit matrix, as the data is already provided as a full sky map on the resolution we desire.
Here, we decompose the noise $n_\mathrm{ff}$ into two components namely a observed one obtained by \texttt{COMMANDER} and an unobserved one, which contains systematics. As we do not consider the correlation structure of the noise in for any data set, the procedure of noise estimation is the same from here on as in \ref{subsubsec:the_noise_model_faraday}, apart from different hyper priors. In this case, we choose $\alpha_{\mathrm{ff}} = 1$ and $\beta_{\mathrm{ff}} $ such that the expectation value over $\ln(\eta_{\mathrm{ff}} )$ is zero. This is Jeffreys prior for the inverse gamma distribution \citep{OLD_FARADAY_2012A&A...542A..93O}.
The lower branches of the free free sky model are shown in Fig. \ref{fig:sky_model_free_free}. The full extended model for the hierarchical Bayesian model is shown in Fig. \ref{fig:sky_model_complete}. Using the considerations above, the likelihood for the free-free sky is again Gaussian in analogy to Eq. \eqref{eq:faraday_likelihood}
\begin{equation}
\label{eq:free_free_likelihood}
\mathcal{P}\left(d_{\mathrm{ff}} \right) = \mathcal{G}\left(d_{\mathrm{ff}}  - \mathcal{R}_{\mathrm{ff}} \mathrm{EM}_\mathrm{ff}|\widetilde{N}_{\mathrm{ff}} \right),
\end{equation}
with the noise covariance $\widetilde{N}_\mathrm{ff}$ defined in analogy to Eq. \eqref{eq:faraday_effective_noise}. 
Again, we still need to explain the detailed model for $\mathrm{EM}_\mathrm{ff}$ and its priors. This is presented in the next section.

\subsection{The free free model}
\label{subsec:the_model_free_free}

\begin{figure*}
\begin{subfigure}{0.49\linewidth}
\centering
\includegraphics[width=1.\textwidth]{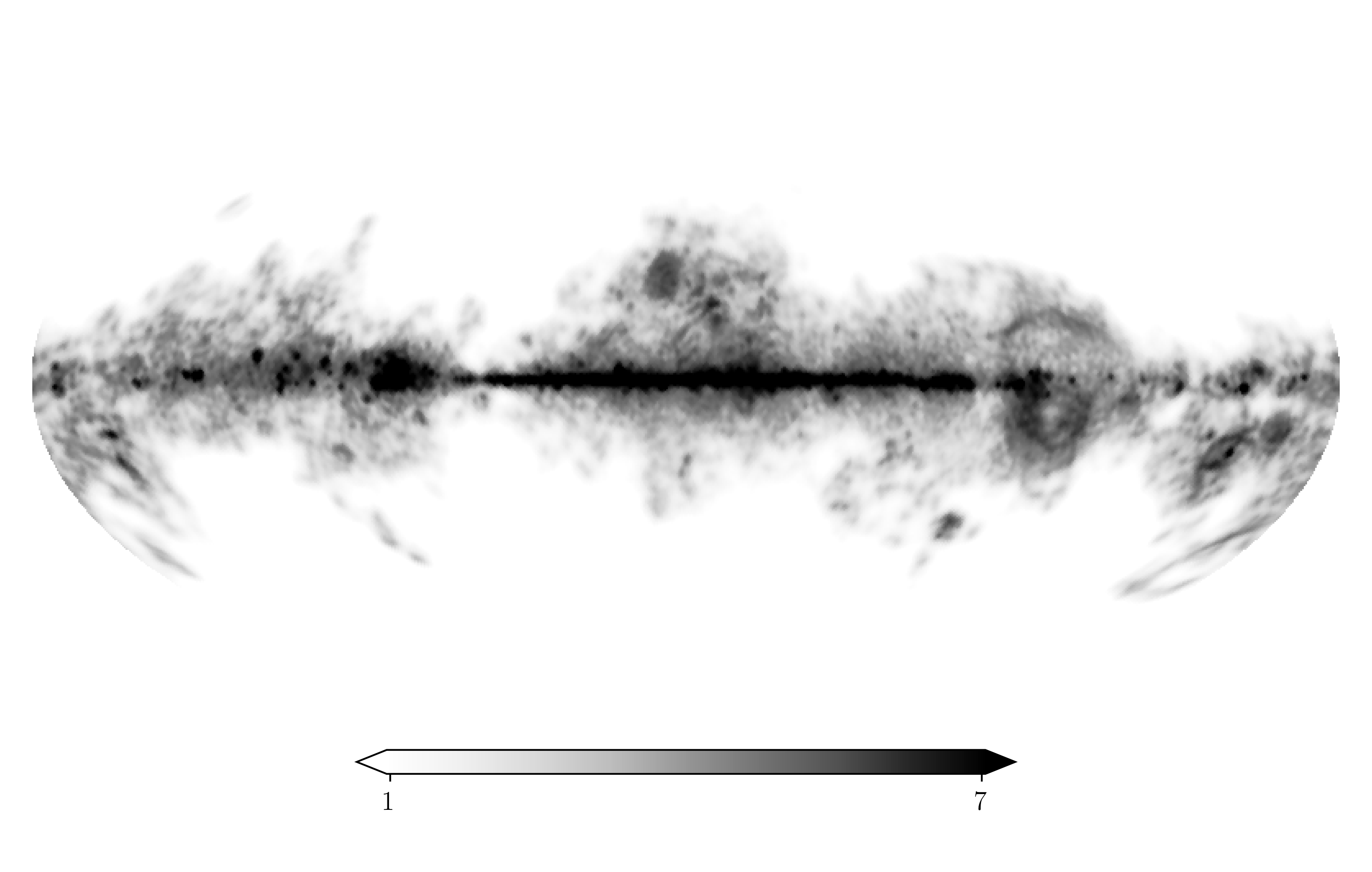}
\caption{\label{fig:rev_epslion}}
\end{subfigure}
\begin{subfigure}{0.49\linewidth}
\centering
\includegraphics[width=1.\textwidth]{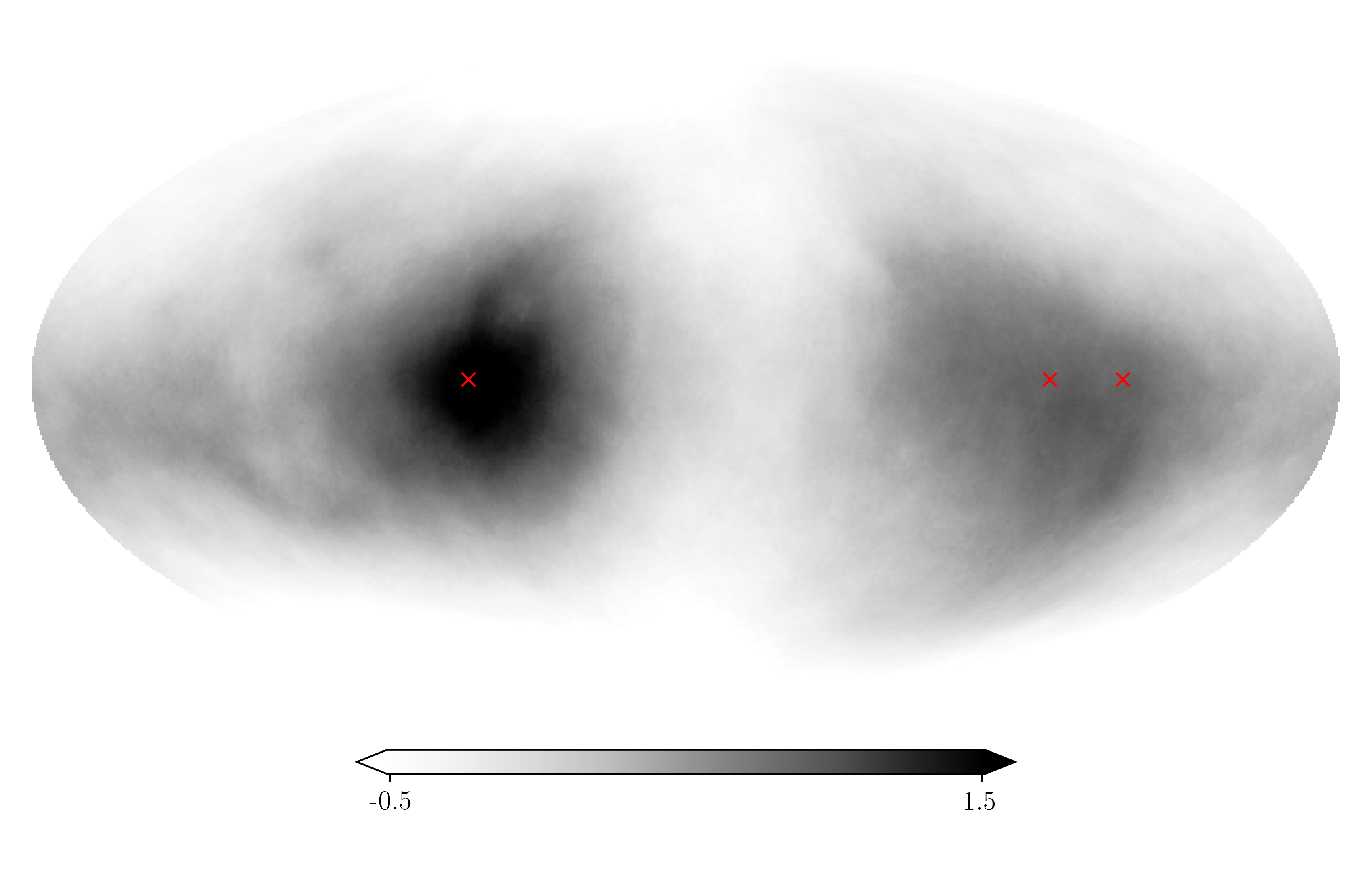}
\caption{\label{fig:rev_delta}}
\end{subfigure}
\begin{subfigure}{0.49\linewidth}
\centering
\includegraphics[width=1.\textwidth]{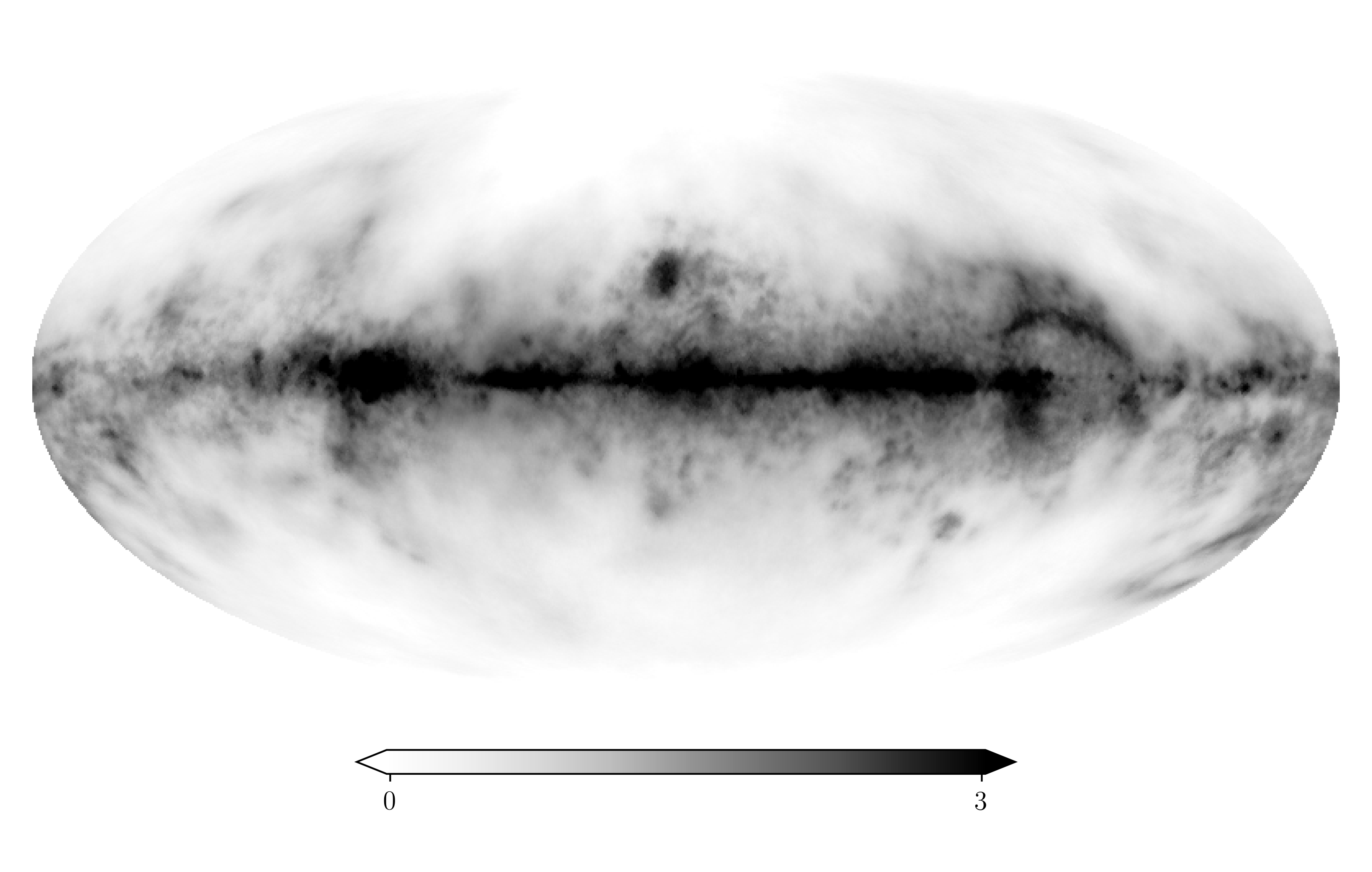}
\caption{\label{fig:rev_log_amplitude_no_rho}}
\end{subfigure}
\begin{subfigure}{0.49\linewidth}
\centering
\includegraphics[width=1.\textwidth]{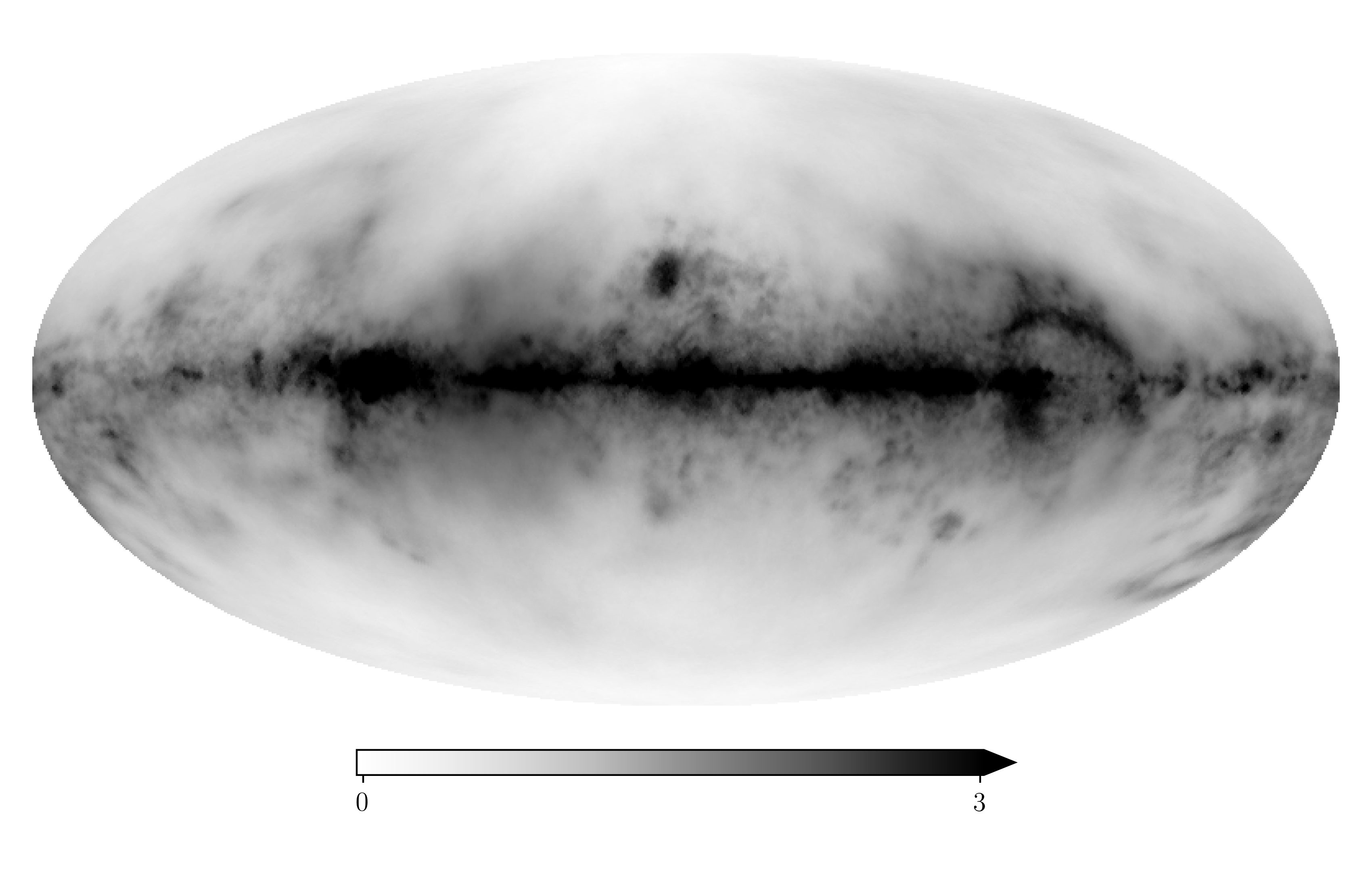}
\caption{\label{fig:rev_log_amplitude}}
\end{subfigure}
\caption{\label{fig:log_amplitudes} Logarithmic amplitude fields. Figure (a) shows the field $\epsilon$ defined in Eq. \eqref{eq:new_faraday_model}, constraint by free-free and Faraday data. This is also the reconstructed log free free map. Figure (b) shows the second amplitude field $\delta$ defined in Eq. \eqref{eq:new_faraday_model}. The red crosses indicate the approximate angular positions of the Orion arm, as given by \citet{Orion_outer_2008ApJ...672..930V} and \citet{Orion_inner_2009ApJ...693..413X}. Figure (c) shows the logarithmic amplitude of the Faraday sky as defined in Eq. \eqref{eq:new_faraday_model_lin} without the additional $\delta$ contribution, while this contribution is included in figure (d).}
\end{figure*}

\begin{figure}
\begin{subfigure}{\linewidth}
\centering
\includegraphics[width=1\textwidth]{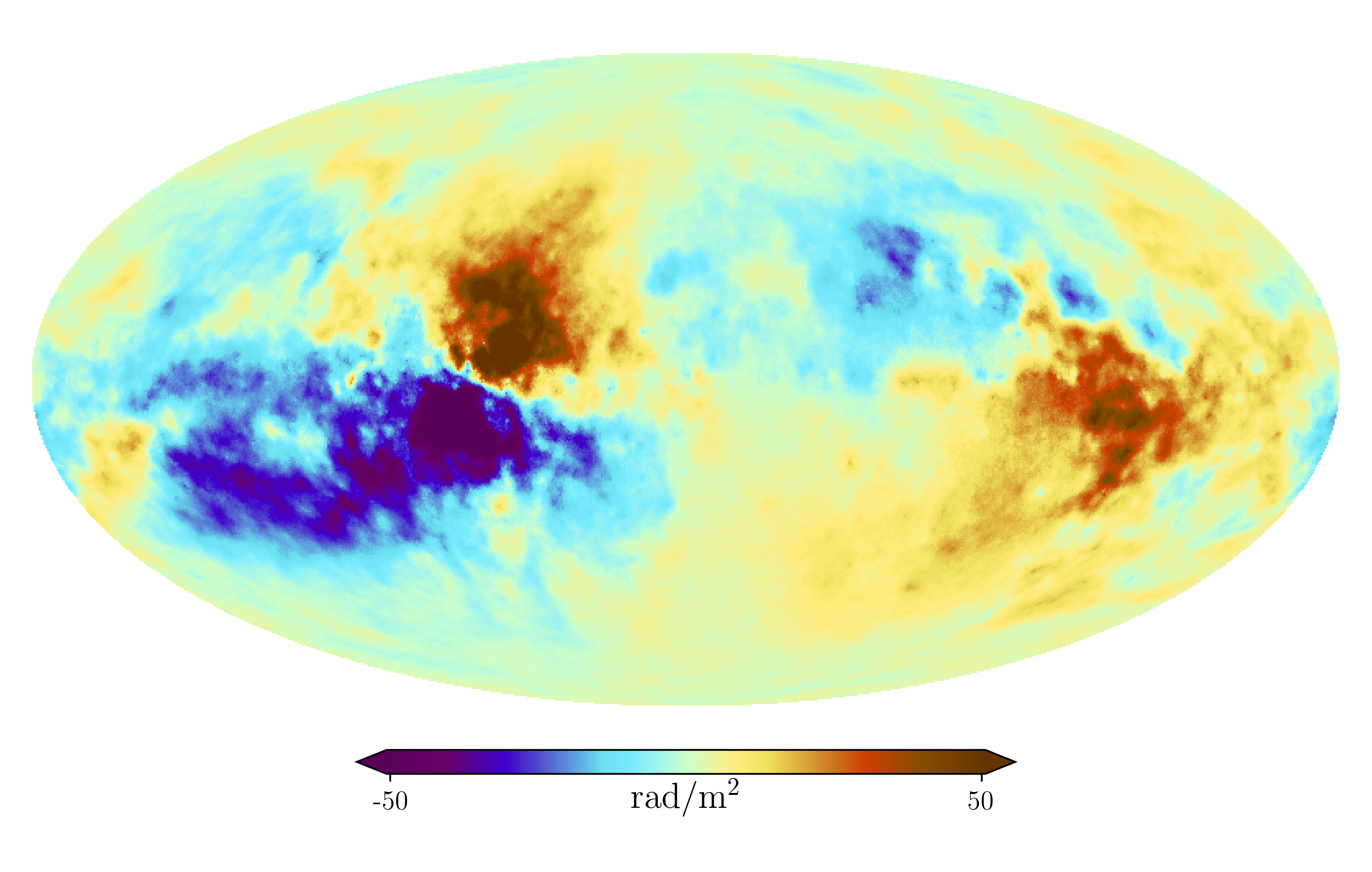}
\caption{\label{fig:rev_faraday_orion}}
\end{subfigure}
\begin{subfigure}{\linewidth}
\centering
\includegraphics[width=1\textwidth]{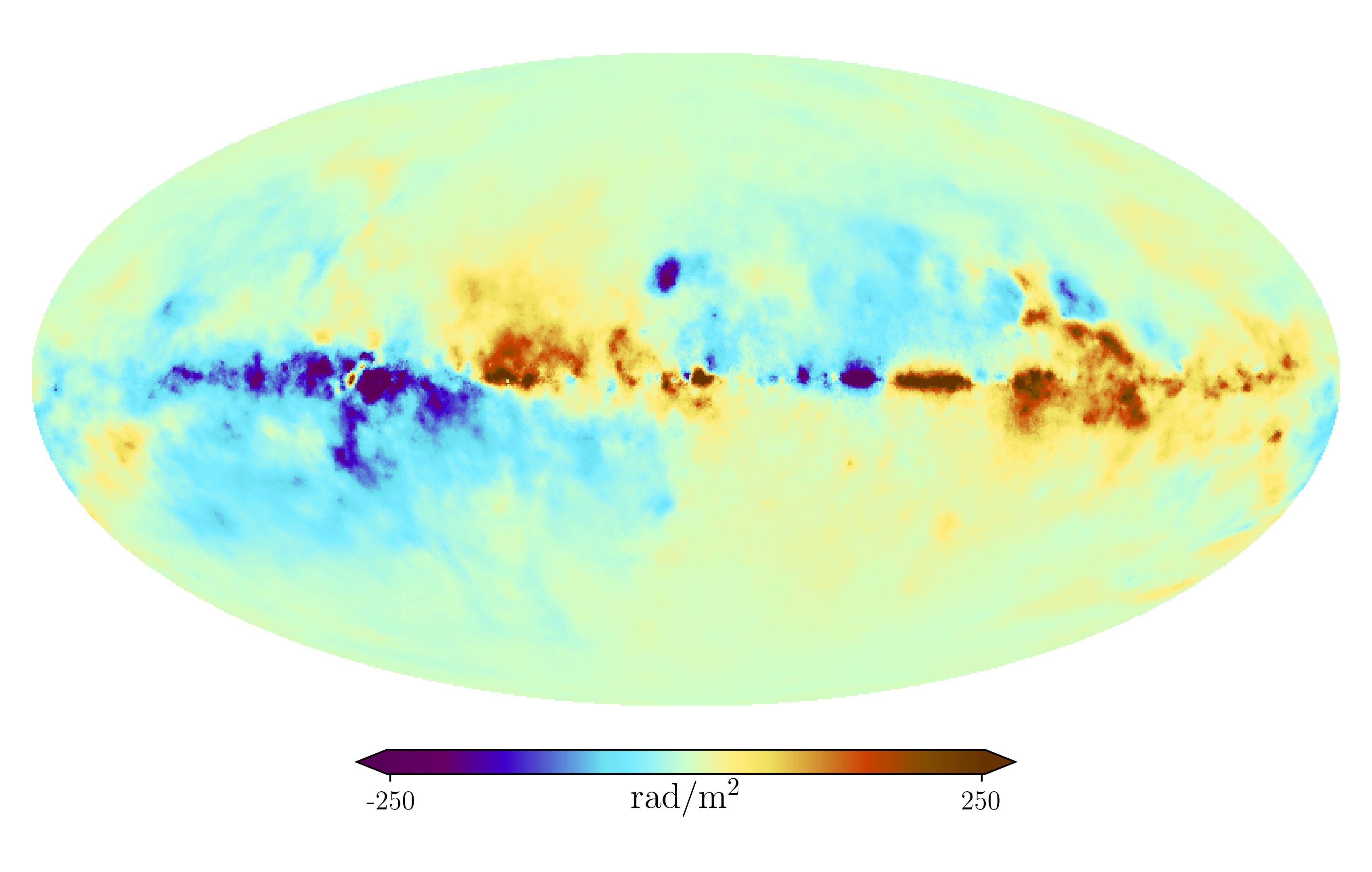}
\caption{\label{fig:rev_faraday_no_orion}}
\end{subfigure}
\caption{\label{fig:faraday_orion} Faraday sky containing (a) only the amplitude contribution from $\delta$ and (b) only the other contributions (see Fig. \ref{fig:rev_delta}). Figure (a) is described by Eq. \eqref{eq:orion}.}
\end{figure}

\subsubsection{The sky model}
\label{subsubsec:the_sky_model_free_free}
We now turn our head to the modeling of the free-free sky and its connection to the amplitude field of the Faraday sky under the constraint of the physical limitations considered in  Sec. \ref{subsec:the_physics_free_free}. We model $\mathrm{EM}_\mathrm{ff}$ with a log-normal approach, again to enforce positivity and to capture the expected large variability of $\mathrm{EM}_\mathrm{ff}$ on the sky:
\begin{equation}
\label{eq:free_free_model}
\mathrm{EM}_\mathrm{ff} \equiv  e^{\epsilon}\cdot \frac{\mathrm{pc}}{\mathrm{cm^6}}
\end{equation}
If the true free-free sky would be a perfect estimator for the amplitude field of the Faraday sky, we could simply replace $\rho$ in Eq. \eqref{eq:faraday_model} with $\epsilon$ to simultaneously infer both quantities from both data sets. However, as discussed in Section \ref{subsec:the_physics_free_free} a necessary condition for a combination of the information of free-free EM's and Faraday rotation measures is to account for magnetic field reversals, different scalings with the thermal electron density and potential systematics in the data sets. All issues can be modeled by introducing further terms to the model.\\

We start with our approach to model the differences in the electron density scaling and the geometry factor introduced by the sign reversals. As we can only work with the LOS integrated quantities, the most natural way of inferring the three dimensional effective scaling factor $\frac{1+p_B}{2}$  in Eq. \eqref{eq:free_free_faraday} is not possible, as this would necessarily imply setting up the full 3D problem, for which we are lacking the necessary radial information. 
We therefore choose to introduce two scaling fields $\gamma: S^2 \rightarrow \mathbb{R}$ and  $\psi: S^2 \rightarrow \mathbb{R}$ into the modeling according to Eq. \eqref{eq:new_faraday_model_lin} which essentially sets up a linear fit for the logarithm of the Faraday amplitude field in terms of the reconstructed logarithmic free-free sky:
\begin{align}
\label{eq:new_faraday_model_lin}
\phi &= \left( e^{\gamma\epsilon + \psi }\right) \chi.
\end{align}
This model somewhat mimics the approximation derived in Eq. \eqref{eq:field_geometry}, with $\gamma$ approximating the role of the power law coefficients and vice versa $e^\psi$ the geometry factor.
Unfortunately, we loose the interpretation of the $\gamma$ field in terms of $\frac{1+p_B}{2}$ in  Eq.\eqref{eq:free_free_faraday} due to the aforementioned lack of consistent 3D modeling. Furthermore the large scale modes of the $\psi$ and $\gamma$ fields are to a degree degenerated, making a quantitative interpretation of the $\psi$ field very hard, although the smaller scaled morphology may be informative. 
Altogether, these terms should be able to capture sign reversal effects as well as strong deviations from the proportionality of the magnetic field strength on the thermal electron density presented in Eq. \eqref{eq:free_free_faraday}. \\
We further have to consider the case where the Faraday sky contains amplitude structure not present in the $\mathrm{EM}_\mathrm{ff}$ sky. This is done by expanding the model according to
\begin{align}
\label{eq:new_faraday_model}
\phi &= \left( e^{\gamma\epsilon + \psi } + e^\delta \right) \chi.
\end{align}
Here, the second amplitude field $\delta: S^2 \rightarrow \mathbb{R}$  is not connected to the free-free map, giving it the possibility to capture structures that are in the Faraday data but not in the free-free data.  
All newly introduced fields are assumed to follow Gaussian statistics, so their prior is similar to Eq. \eqref{eq:gaussian_prior}, again with unknown correlation structure. One should note at this point that the obvious a priori degeneracies between the newly introduced helper fields are usually broken during the inference, if sufficiently supported by data. This is possible due to the joint correlation structure inference, which constrains the respective fields just to the degree it is demanded by the inference problem.\\
Although there is motivation for each term, the above modeling might be viewed as ad hoc, as one may certainly argue for the addition or omission of certain terms or a completely different modeling altogether. There are certain limitations for potential models, as they all have to be able to capture the physical characteristics (say e.g. the sign reversal of the Faraday map), and should be kept as simple as possible. Unfortunately these requirements already allow for a range of different parametrizations.  We have varied the model during our analysis by e.g. adding or omitting further terms in Eq. \eqref{eq:new_faraday_model}. We found little variation in the overall morphology of the Faraday sky, apart form the disk, where e.g omission of the $\gamma$ and  $\psi$  field lead to an extremely pronounced features, with RM values reaching regimes a magnitude higher than previous reconstructions.
As we think that the above presented physical considerations are too important to be neglected, we deem the resulting maps in these cases as unrealistic. 
Given that all newly introduced helper fields have captured distinct structures as we will show in Sec. \ref{subsec:the_results_free_free}, we view the model in Eq. \eqref{eq:new_faraday_model} as one of the simplest cases which can capture all relevant effects, but nonetheless we deem the systematic errors higher than the statistical ones in this inference.\\
To gain some insights on the robustness of the obtained free-free map, we will further perform an inference just for the free free sky with the model in Eq. \eqref{eq:free_free_model}, unconstrained by Faraday data. The comparison of the result to the outcome of the joint Faraday and free-free inference will be presented in the results section. We can implement this extended model numerically with similar algorithms as described in Sec. \ref{subsec:the_inference_faraday}.

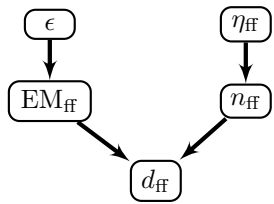
\begin{figure}
\centering
% \begin{center}
\begin{tikzpicture}[auto]
%    \small
    % node placement with matrix library
  \matrix[ampersand replacement=\&, row sep=0.5cm, column sep=0.5cm] {
    % Place nodes
    \&
    \&
    \node [block] (delta) {$\epsilon$};
    \&
    \&
    \node [block] (eta) {$\eta_\mathrm{ff}$};   
    \&
    \&
    \\
    \&
    \&
    \node [block2] (j) {$\mathrm{EM}_\mathrm{ff}$};
    \&
    \&
    \node [block] (n) {$n_\mathrm{ff}$};
    \&
    \& 
    \\
    \&
    \&
    \&
    \node [block] (d_j) {$d_\mathrm{ff}$};
    \&
    \&
    \&
    \\ 
    }; 
    % Draw edges
    \path [line] (delta) -- node{}(j);
    \path [line] (j) -- node{}(d_j);
    \path [line] (n) -- node{}(d_j);
    \path [line] (eta) -- node{}(n);

\end{tikzpicture}

% \end{center}
\caption{\label{fig:sky_model_free_free} Hierarchical tree model for the free free EM sky. We model $\mathrm{EM}_\mathrm{ff}$ via the exponentiated field  $\epsilon$ (see Eq. \eqref{eq:free_free_model}). The field $\mathrm{EM}_\mathrm{ff}$ is together with the noise $n_\mathrm{ff}$ connected to the observed data via Eq. \eqref{eq:free_free_data}}.
\end{figure}   

\begin{figure*}
\centering
% \makeatletter\onecolumngrid@push\makeatother
% \begin{center}
\begin{tikzpicture}[auto]
%    \small
    % node placement with matrix library
  \matrix[ampersand replacement=\&, row sep=0.5cm, column sep=0.5cm] {
    % Place nodes
    \node [block] (eta_ff) {$\eta_\mathrm{ff}$};
    \&
    \&
    \&
    \node [block] (rho) {$\epsilon$};
    \&
    \node [block] (gamma) {$\gamma$};    
    \&
    \node [block] (chi) {$\chi$};
    \&
    \node [block] (psi) {$\psi$};
    \&
    \node [block] (delta) {$\delta$};
    \&
    \node [block] (eta_phi) {$\eta_\phi$};
    \\  % newline
    \node [block] (n_ff) {$n_{\mathrm{ff}}$};
	\&
    \&
    \node [block2] (ff) {$\mathrm{EM}_\mathrm{ff}$};
    \&
    \&
    \&
    \node [block] (phi) {$\phi$};
    \& 
    \&
    \&
    \node [block] (n_phi) {$n_\phi$};
    \&
    \\ % newline
    \&
    \node [block] (data_ff) {$d_\mathrm{ff}$};  
    \&
    \&
    \&
    \&
    \&
    \node [block] (data_phi) {$d_\phi$};  
    \&
    \&\\
    }; 
    % Draw edges
    \path [line] (phi) -- node{}(data_phi);
    \path [line] (n_phi) -- node{}(data_phi);
    \path [line] (ff) -- node{}(data_ff);
    \path [line] (n_ff) -- node{}(data_ff);

    \path [line] (eta_phi) -- node{}(n_phi);
    \path [line] (eta_ff) -- node{}(n_ff);
    
    \path [line] (rho) -- node{}(ff);
    \path [line] (rho) -- node{}(phi);
    \path [line] (gamma) -- node{}(phi);
    \path [line] (chi) -- node{}(phi);
    \path [line] (psi) -- node{}(phi);
    \path [line] (delta) -- node{}(phi);

    % \path [curved line] (alpha_beta) to node{}(eta);

\end{tikzpicture}
% \end{center}
% \makeatletter\onecolumngrid@pop\makeatother
\caption{\label{fig:sky_model_complete} The full hierarchical model excluding the power spectrum hyper priors for the fields. The lowest layer contains the data sets $d_\mathrm{ff}$ and $d_\phi$. These are connected by the equations \eqref{eq:faraday_data} and \eqref{eq:free_free_data} to the sky maps $\mathrm{EM}_\mathrm{ff}$ and $\phi$. The sky maps in turn are connected to the Gaussian fields in the uppermost layer via the respective models defined in Eqs.  \eqref{eq:free_free_model} and \eqref{eq:new_faraday_model}. From there on, the higher branches for the respective correlation structure inference follow. These are not depicted here, but shown and explained in the Appendix.}
\end{figure*}
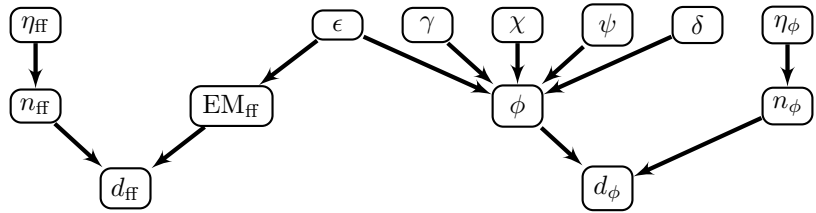

\subsection{The Results}
\label{subsec:the_results_free_free}

\subsubsection{The revised Faraday map}
In the following we show the results of the model depicted in Fig. \ref{fig:sky_model_complete}. 
We show the mean, the differences to the previous and initially revised reconstructions and the power spectrum in Figs. \ref{fig:rev_faraday_mean}, \ref{fig:diff_pre_rev}, \ref{fig:diff_rev_ini}  and \ref{fig:power}, respectively. The inclusion of the free-free sky has lead to a notable difference in our newly revised estimate of the Faraday sky. Driven by the strong disc in the free-free data, the Faraday sky now has a similarly strongly pronounced disc feature as well. Other features mostly farther away from the Galactic plane remained relatively stable, such as the northern arc of the Gum nebula, which is also strongly visible in the free-free data. 
In Fig. \ref{fig:rev_faraday_var}, we show the updated uncertainties. These have again narrowed considerably compared to our initial free-free data free reconstruction as shown in Fig. \ref{fig:ini_faraday_var}. The innermost part of the disc, however, is still quite uncertain. Furthermore one should be aware that any model uncertainties are not considered in this plot.
The new power spectrum of the Faraday map is shown as the red line in Fig. \ref{fig:power}. It is very similar to that of the previous reconstruction, with an notable offset towards smaller scales, implying a slightly steeper power law. 

\begin{figure}
\begin{minipage}{\linewidth}
\centering
\includegraphics[width=1\textwidth]{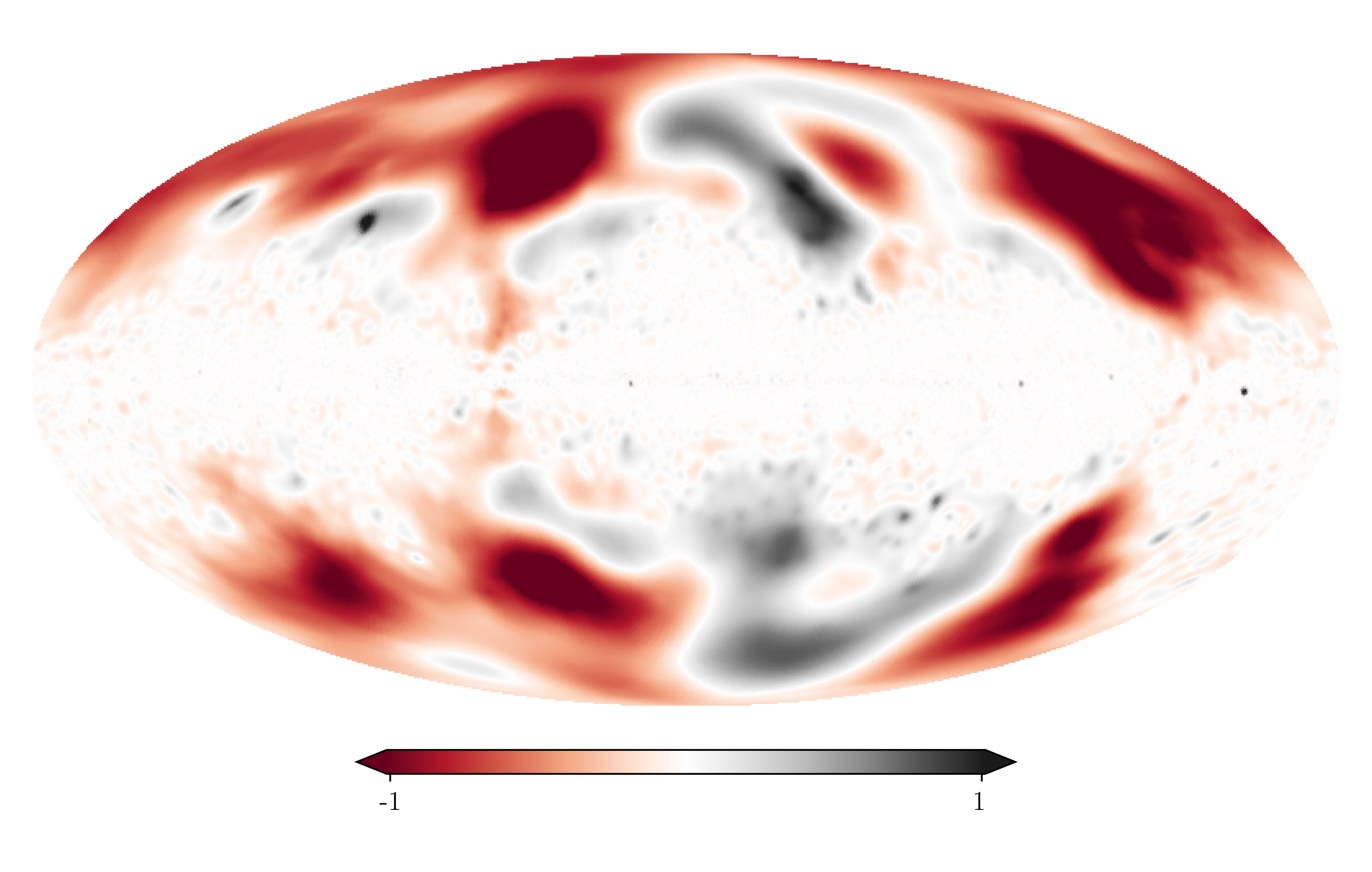}
\caption{\label{fig:diff_epsilon} Difference between the field $\epsilon$ defined in Eq. \eqref{eq:free_free_model}, inferred only by free-free data and the same field resulting from the joint inference with Faraday data.}
\end{minipage}
\end{figure}
\begin{figure}
\begin{minipage}{\linewidth}
\includegraphics[width=1\textwidth]{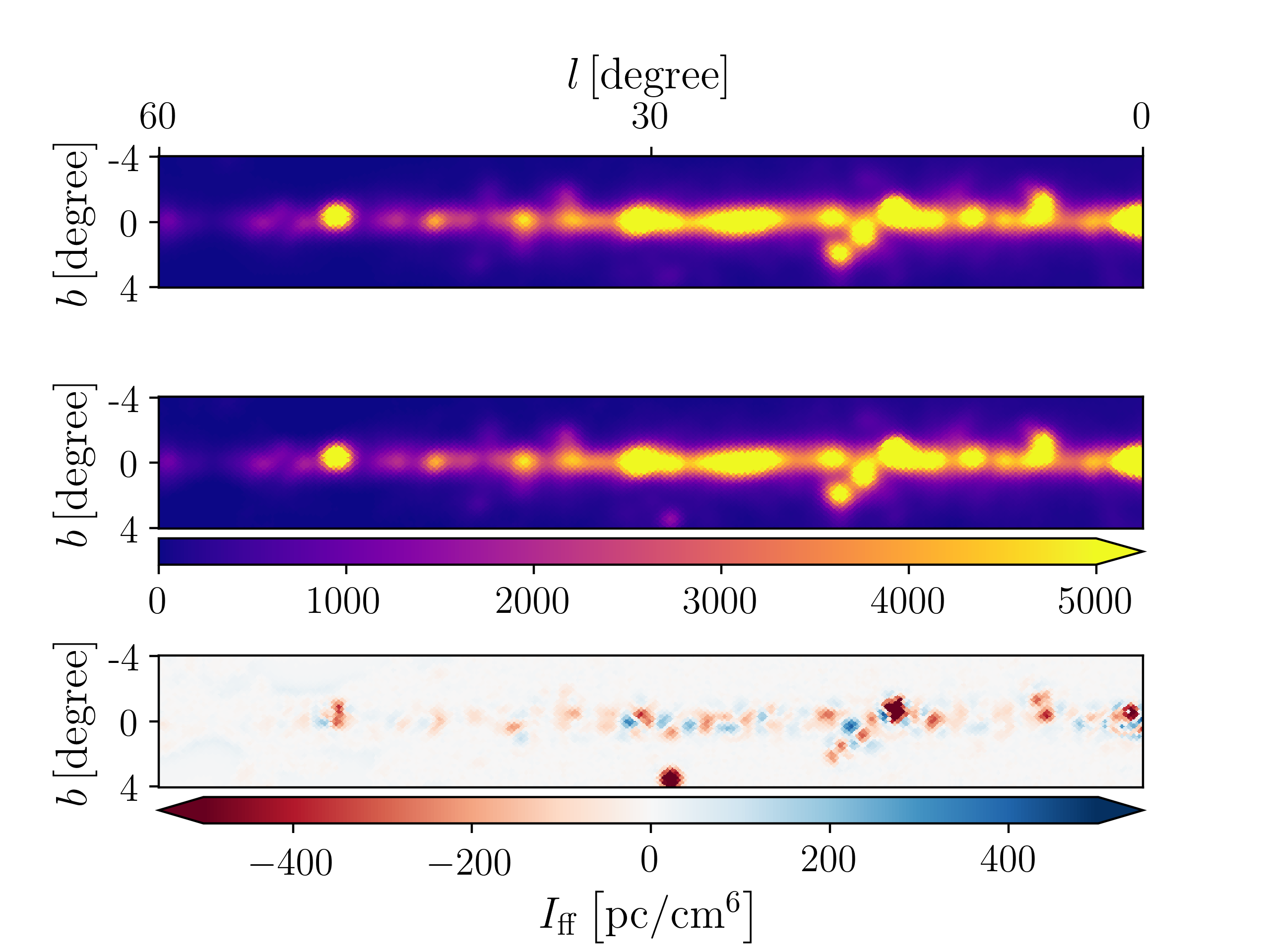}
\caption{\label{fig:planck_diff_cut} Excerpt of the free-free reconstruction in the first slice, followed by the same excerpt in the Planck data. The difference of the two slices is shown in the last row.}
\end{minipage}
\end{figure}

\subsubsection{The components}
We now show some of the components of the Faraday model in Eq. \eqref{eq:new_faraday_model}, apart from the $\epsilon$ field in Fig. \ref{fig:rev_epslion}, which will be more closely discussed in the following section dedicated to the inference results of the free-free sky. \\
The revised Faraday amplitude is shown in Fig. \ref{fig:rev_amplitude}. The comparison to the initially revised amplitude field in Fig. \ref{fig:ini_amplitude} and the free-free data in Fig. \ref{fig:planck} demonstrated the influence of both the Faraday and the free-free data on the new field. The field is enhanced by roughly a factor of 4 compared to the old amplitude field. 
This may be a result of the symmetry breaking process described in Sec. \ref{subsec:the_model_faraday}, as now the amplitude field is partially constrained by the free-free sky. The comparison of the sign fields will demonstrate that this factor was mostly absorbed there.\\
The second amplitude field $\delta$ shown in Fig. \ref{fig:rev_delta}, its influence on the total logarithmic Faraday amplitude is demonstrated by Figs. \ref{fig:rev_log_amplitude_no_rho} and \ref{fig:rev_log_amplitude}. The field is dominated by two diffuse structures of enhanced $\delta$, centered in the Galactic plane and separated by roughly \ang{180} longitude. Remembering the discussion in Sec. \ref{subsec:the_physics_free_free}, the $\delta$ field is supposed predominately capture amplitude variations between the free-free and the Faraday sky. Larger values of $\delta$ can be caused by systematic errors in either the free-free or the Faraday data and/or a high alignment of the magnetic field with the LOS in this direction. The approximated longitudinal difference of the two $\delta$ enhancements of about \ang{180} as well as their positions seem to correspond to the angular positions of the Orion arm on the sky, which are depicted in the plot as red crosses. In the first quadrant of the Milky Way, at about \ang{60} longitude, this minor arm is known to extend over several kiloparsec up to the Sagittarius arm, with which it might merge at last \citep{Orion_inner_2009ApJ...693..413X}. In the other direction towards the third quadrant of the Galaxy, the structure of the arm is much more complex. Even a bifurcation or a crossing with the Perseus arm seems possible \citep{Orion_outer_2008ApJ...672..930V}. We have indicated possible continuations of the arm in the plot at \ang{240} and \ang{260} longitude, respectively. The $\delta$ field seems to reflect not only the position of the arm, but also its morphology, as the structure in the third quadrant is more diffuse, as opposed to the relatively pronounced structure at \ang{60} longitude.
All in all, this would argue for a high magnetic field alignment as a viable possibility for the interpretation of the second amplitude field. As the Sun is positioned directly in the Orion arm, we expect more coherent magnetic fields along the LOS directed pointed directly to the dominant direction of the arm. Remembering the discussion in Sec. \ref{subsec:the_physics_free_free}, this then may lead to a disproportionate increase in the observed Faraday measures as the free-free map would indicate at these positions. Furthermore the Faraday sky in Fig. \ref{fig:rev_faraday_mean} indeed shows opposite magnetic signs for those regions, as one would expect for a field traversing the solar location. We show the field 
\begin{equation}
\label{eq:orion}
\phi_\mathrm{Orion} = e^\delta\chi
\end{equation}
in Fig. \ref{fig:rev_faraday_orion}, which corresponds to our estimate of the enhancement of the Orion Faraday depth, should the above considerations be true and if the $\chi$ field really only captured the sign. The (in terms of amplitude contributions) complementary Faraday component is shown in Fig. \ref{fig:rev_faraday_no_orion}. The considerable differences to Fig. \ref{fig:rev_faraday_mean} in the sky regions corresponding to the Orion arm reveal again the influence of the $\delta$ field there.   
Of course we can not rule out any of the other explanations for the observed $\delta$ enhancements, the plot in Fig. \ref{fig:rev_faraday_orion} should therefore be taken with care.\\
The comparison of the amplitude field in Fig. \ref{fig:rev_log_amplitude_no_rho} without the $\delta$ contributions with the $\epsilon$ field alone (see Fig. \ref{fig:rev_epslion}) reveal the strong rescaling and reweighting of the free-free contribution due to the $\gamma$ and $\psi$ fields. \\
At last we discuss the sign field $\chi$ in Fig. \ref{fig:rev_sign}. Little has changed in this component morphologically compared to Fig. \ref{fig:ini_sign}. The absolute value of the field has decreased by approximately a factor of 4. This factor was mostly absorbed into the new amplitude field, as discussed before. Apart from that, the field has again fully served its intended purpose, namely capturing the sign of the Faraday sky.\\
We now discuss the results of the inference of the free-free EM sky.

\subsubsection{The revised free-free map}

The revised free-free EM sky constrained by both the Planck and the Faraday data is shown in Fig. \ref{fig:rev_free_free}, the logarithm of it is shown in Fig. \ref{fig:rev_epslion}. 
The difference to the map inferred using only Planck data is shown in Fig. \ref{fig:diff_epsilon}. There is little difference between the maps, indicating that the systematic differences of the data sets were mostly absorbed by the second amplitude field $\delta$ and the rescaling fields $\gamma$ and $\epsilon$. Notable differences appear only towards the Galactic poles, where Planck has measured little to no free-free component, while considerable Faraday rotation exists. 
The visual comparison of Fig. \ref{fig:rev_free_free} to the data shown in Fig. \ref{fig:planck} demonstrates good accordance. The main difference is the lack of small structures far away from the disc, which where mostly absorbed into the noise.
Closer inspection reveals some small scaled deviations in the disc, which we want to investigate by showing an excerpt of the disc ranging from \ang{0} to \ang{60} longitude in Fig. \ref{fig:planck_diff_cut}. This shows approximately the same region as the first plot of Fig. 7 in \citet{LOW_DIFFUSE_2016A&A...594A..25P}, where the Planck team compares their free-free temperature map with independent measurements via the radio recombination line survey (RRL). While our reconstruction reproduced the overall morphology, some peaks in the picture experience deviations around $20\,\%$. In general our results seem to underestimate the amplitudes compared to Planck. It is noteworthy that the comparison between Planck and RRL resulted in a similar statement, as the Planck features were generally much more pronounced.
We will refrain from drawing physical conclusions from these discrepancies as these can only be resolved by a detailed discussion of the Planck data analysis, which is beyond the scope of this work. Note again that the Faraday data has little influence on this result. The changes are mainly driven by the noise estimation and the full consideration of the correlation structure of the free-free EM sky. We think that this deems the inclusion of the correlation structure in the inference an important step for any component separation algorithm, as also shown in \citet{COMP_JAKOB_2017PhRvE..96d2114K}.

\section{Summary and conclusion}
\label{sec:conclusion}
We infer the Galactic Faraday sky map using the same data as \citet{OLD_FARADAY_2012A&A...542A..93O} using an updated algorithm. This involves the separation of the Faraday sky into the point wise product of a sign and amplitude field, which both are supposed to mainly capture the sign and the Galactic profile of the Faraday sky, respectively. In the first part of the paper, we can largely confirm the old results in map and power spectrum. Smaller differences can be explained by the algorithmic advancements. The biggest improvement here is the updated uncertainty map, where we were able to significantly reduce the error bars over the whole sky, especially in some locations in the galactic disc.\\
In the further proceedings we notice that the amplitude field has notable similarities to the Galactic free-free EM map. In the second part of the paper we therefore incorporate free free data as a potential proxy for the Faraday amplitude field into our inference.
This leads to a new Faraday map which now is notably different to previous results. The Galactic disc is much more pronounced and uncertainties are largely reduced. Since a pronounced Galactic is very sensible, as well as required by the data, which now is more satisfied as in any of the other approaches, we regard this map as the most reliable and therefore to be the main result of this work. Moreover, the component fields introduced in our model to resolve discrepancies between the two data sets reveal indicators for more LOS aligned magnetic field structures in the direction of the Orion arm.
We also produce a denoised free-free map, which agrees with the Planck data to large extent. 

\begin{acknowledgements}
	We acknowledge fruitful discussions with J. Knollm\"uller and others from the information field theory group at the MPI for Astrophysics, Garching.
\end{acknowledgements}

\clearpage
\bibliographystyle{aa}
\bibliography{lib}

\begin{appendix}
\label{sec:appendix_a}

\section{Generic IFT concepts}
\label{sec:generic_ift_concepts}

These appendices are intended to provide the details on the inference scheme used in this work. We will start with introducing some core concepts of IFT. After that, the details of the correlation structure inference are explained. In this part, we will also show the equations for the full problem posterior of our model above. We close with some details on the used posterior evaluation scheme. Extended and more mathematically rigorous discussions on these topics can be found in \citet{OLD_IFT_PhRvD..80j5005E, JAKOB_2017arXiv171102955K, IFT_2018arXiv180403350E, JAKOB_2_2019arXiv190111033K}. \\
A generic expression for the posterior probability distribution for a inference problem can be written using Bayes theorem:

\begin{equation}
\label{eq:generic_bayes}
\mathcal{P}\left(s\vert d\right) = \frac{\mathcal{P}\left(d| s\right)\mathcal{P}\left(s\right)}{\mathcal{P}\left(d\right)} 
\end{equation}

In our case, the signal $s$ contains all fields that describe the aforementioned properties of the desired sky maps. The left hand sight of Eq. \eqref{eq:generic_bayes} is commonly called the posterior. The right hand side consists of the prior $\mathcal{P}\left(s\right)$, the likelihood $\mathcal{P}\left(d| s\right)$ and the evidence $\mathcal{P}\left(d\right)$. The latter is important for model comparison tests, but negligible in an inference algorithm that aims to find the optimal field realizations of $s$ for a given model, as it does not depend on $s$, but only serves as a normalization constant. 

In the further proceeding, we will consider the information Hamiltonian or negative log-probabilities of the above equation,

\begin{equation}
\label{eq:hamiltonian}
\mathcal{H}\left(s\vert d\right) = -\ln \mathcal{P}\left(s\vert d\right)
\end{equation}

 This has mostly convenience reasons, as the resulting sums of log-probabilities are usually numerically better behaved as the product of probabilities. Furthermore there is a nice analogy to statistical mechanics and field theory (hence the use of `Hamiltonian'), which results in a rich variety of techniques IFT can make use of.\\

Already using the expressions for the likelihoods from Eqs. \eqref{eq:faraday_likelihood} and \eqref{eq:free_free_likelihood}, the respective Gaussian priors similar to Eq. \eqref{eq:gaussian_prior} and the terms stemming from the noise estimation models in Eq. \eqref{eq:noise_model}, we can write down the posterior Hamiltonian assuming independent measurements of $\mathbf{d} = \left(d_\phi, d_{\mathrm{ff}}\right)^T$:

\begin{align}
\label{eq:generic_hamiltonian}
& \mathcal{H}\left(\mathbf{s}\vert \mathbf{d}\right) \equiv -\log\left(\mathcal{P}\left(\mathbf{s} \vert \mathbf{d}\right)\right) \equalhat \nonumber\\
& \sum_{d \in \mathbf{d}}  \mathcal{H}\left( d\vert \mathbf{s}\right) + \sum_{f \in \mathbf{s}}\mathcal{H}\left(f\right) \equalhat \nonumber \\
& \sum_{d \in \mathbf{d}} \frac{1}{2}\left(d-\mathcal{R}_d \mathbf{s}\right)\widetilde{N}_d^{-1}\left(d-\mathcal{R}_d \mathbf{s}\right)^\dagger + \nonumber\\
& + \frac{\beta_d}{\eta_d} + (\alpha_d + 1)\ln(\eta_d) +  \sum_{f \in \mathbf{s}} f^\dagger S^{-1}_ff.
\end{align}
The vector $\mathbf{s} \equiv \left(... \right)^T$ contains all Gaussian random fields introduced in the previous sections, which are assumed to be a priori independent from each other. The symbol $\equalhat$ indicates the omission of unnecessary additive constants.
This equation already contains most of the model. A crucial part is missing, however, namely we have not specified the prior covariances $S_f$. The modeling and inference of those is explained in the next section.

\section{Correlation structure model}
\label{sec:correlation_structure_model}

Again we will not follow the details in \citet{OLD_FARADAY_2012A&A...542A..93O}, but use an updated algorithm for the correlation structure inference. A short discussion on the differences of the two approaches and potential implications on the spectrum follows at the end of this section. We have verified our results via a comparison to the previous map only using the data used in \citet{OLD_FARADAY_2012A&A...542A..93O} in Sec. \ref{subsec:results_faraday}.\\
The assumption of statistical homogeneity and isotropy together with Gaussianity for a field $f$ implies via the Wiener-Khintchin theorem that the respective covariance is diagonal in harmonic space. 
We will make use of this theorem, as it allows us to describe the correlation structure via a field with at most $N_{\mathrm{pix}}$ dimensions, as opposed to the  $N^2_{\mathrm{pix}}$ dimensions in the non-harmonic space.
The correlation structure of a Gaussian field $f$ is then fully described via its power spectrum defined as
\begin{equation}
\label{eq:power_spectrum}
C^{(f)}_{\ell} =    \mathbb{P}_\ell \left\langle \widehat{f} \widehat{f}^*\right\rangle_{\mathcal{P}_{(x)}}=\mathbb{P}_\ell \frac{1}{2\ell+1}\sum\limits_m\vert\widehat{f}_{\ell m}\vert^2.
\end{equation}
where $\mathbb{P}_\ell$ is a projection operator onto spectral bands $\mathbf{\ell}$ of the power spectrum. This operator simply gives us the possibility to reduce the degrees of freedom of the spectrum, it may very well be set to a unit matrix in case one e.g. expects rather sharp features such as characteristic frequencies. \\
For the inference, we follow the reparametrization trick laid out in \citet{JAKOB_2017arXiv171102955K}. We decompose the sky fields into 
\begin{align}
\label{eq:amplitudes}
f = \mathbb{Y}A_{f}\xi_{f}.
\end{align}
$\mathbb{Y}$ is the spherical harmonic transform. The amplitude fields $A_f$ contain the information on the correlation structure on the respective field, while the $\xi_f$ are a priori white Gaussian fields with zero mean and unit covariance which contain the information of the specific field configuration of the field $f$ given the correlation structure in $A_f$. Both fields live in harmonic space. One can show that the amplitude fields are related to the covariance via 

\begin{figure}
\centering
% \begin{center}
\begin{tikzpicture}[auto]
%    \small
    % node placement with matrix library
  \matrix[ampersand replacement=\&, row sep=0.5cm, column sep=0.5cm] {
    % Place nodes
    \node [cloud] (hyper_priors) {$\mathrm{hyper\,priors}$};
    \&
    \node [block2] (sigma_l) {$\sigma_f,\,l_f$};
    \&
    \&
    \node [block] (zeta) {$\zeta_f$};
    \&
    \\
    \node [cloud] (correlations) {$\mathrm{correlations}$};
    \&
    \&
    \node [block] (A) {$A_f$};
    \&
    \&
    \node [block] (xi) {$\xi_f$};
    \\
    \node [cloud] (field) {$\mathrm{field}$};
    \&
    \&
    \&
    \node [block] (f) {$f$};
    \&
    \\ 
    }; 
    % Draw edges
    \path [line] (sigma_l) -- node{}(A);
    \path [line] (zeta) -- node{}(A);
    \path [line] (A) -- node{}(f);
    \path [line] (xi) -- node{}(f);

%    \path [line] (B_rec) -- node {$\mathrm{(g)}$}(B_today);
\end{tikzpicture}

% \end{center}
\caption{\label{fig:field_model} Hierarchical tree model for a generic Gaussian field $f$ living on the sky. It is decomposed into a amplitude field $A_f$ containing the correlation structure and a a priori white Gaussian field $\xi_f$ connecting this correlation information to the actual spatial representation. $A_f$ on its own again depends on hyper priors $l_f,\, \sigma_f $ and another white field $\zeta_f$. This tree wraps up Section \ref{sec:correlation_structure_model} and is the core building block of our correlation structure inference.}
\end{figure}
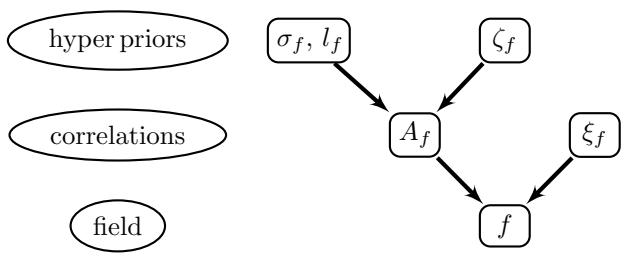

\begin{eqnarray}
\label{eq:amplitude_power}
S_{f} = \mathbb{Y} A_f A_f^\dagger\mathbb{Y}^\dagger.
\end{eqnarray}
This formulation has immediate consequences for the Hamiltonian, as the prior terms are now independent of the correlation structure as 

\begin{equation}
\label{eq:white_prior}
f^\dagger S_f^{-1}f = \xi_f^\dagger \xi_f, 
\end{equation}
using the above equations. For a more general discussion of the advantages of this approach we refer the reader to \citet{JAKOB_2017arXiv171102955K}.
Now we are left with the inference of the amplitude fields. Those are modeled via a log-normal field $\tau_f$ to ensure positivity:

\begin{equation}
\label{eq:amplitude_fields}
A_f = \mathbb{P_\ell}e^{\frac{1}{2}\tau_f}
\end{equation}

The inference then follows the approach of \citet{Arras}, in which $\tau_f$ is again transformed according to the reparametrisation trick above in to a white field $\xi_f$ and hyper parameters $\sigma_f$ and $l_f$ which steer the expected variability in the spectrum. \\
In \citet{OLD_FARADAY_2012A&A...542A..93O} the power spectrum was modeled via a log-normal field together with the product of an inverse gamma and a regularizing smoothness likelihood. The above demonstrated flattening of the hierarchical structure was not performed.\\  
We show a comparison in the section \ref{subsec:results_faraday} to demonstrate that the revised algorithm reproduces the inference of the correlation structure in \citet{OLD_FARADAY_2012A&A...542A..93O}, and also comment on the minor differences occurring between the two spectra. The full Hamiltonian for the simultaneous inference of Faraday sky, the free-free sky, the correlation structure of their respective component fields and noise corrections terms is then:
\begin{align}
\label{eq:full_hamiltonian}
& \mathcal{H}\left(\mathbf{s}|\mathbf{d}\right) \equalhat \nonumber \\
&  \sum_{d \in \mathbf{d}} \frac{1}{2}\left(d-\mathcal{R}_d \mathbf{s}\right)\widetilde{N}_d^{-1}\left(d-\mathcal{R}_d \mathbf{s}\right)^\dagger +  \nonumber \\
& + \frac{\beta_d}{\eta_d} + (\alpha_d + 1)\ln(\eta_d) + \sum_{f \in \mathbf{s}} \xi_f^\dagger \xi_f + \mathcal{H}\left(\tau_f\vert \zeta_f, \sigma_f, l_f\right),
\end{align}
where the last term contains the hyper priors for the $\tau_f$ inference and is described in detail in \citet{Arras}.

\section{Evaluating the posterior} 
\label{sec:evaluating_the_posterior}

The Hamiltonian in Eq. \eqref{eq:full_hamiltonian} fully describes our information on the problem. 
Unfortunately, the non-linearities in this expression imply that this posterior distribution is non-Gaussian, making the evaluation a non-trivial task as a general analytic solution is not known to exist. Therefore, we need to choose an appropriate numerical evaluation scheme. For this end, variational Bayes approaches have proven to combine satisfying accurateness in their uncertainty estimates with speed and stability during the minimization \citep{JAKOB_2017arXiv171102955K}. In a mathematical language, this implies that we will minimize the Kullback-Leibler Divergence (KL)

\begin{align}
\label{eq:KL}
& \mathcal{D}_{\mathcal{KL}}\left(\widetilde{\mathcal{P}}\left(s\vert d\right)\vert\vert \mathcal{P}\left(s\vert d\right)\right) = \nonumber \\ & \int \mathcal{D}s\, \widetilde{\mathcal{P}}\left(s\vert d\right)\ln\left(\frac{\widetilde{\mathcal{P}}\left(s\vert d\right)}{\mathcal{P}\left(s\vert d\right)}\right) = \nonumber \\
& = \left\langle \mathcal{H}\left(s\vert d\right) \right\rangle_{\widetilde{\mathcal{P}}\left(s\vert d\right)} -  \left\langle \widetilde{\mathcal{H}}\left(s\vert d\right) \right\rangle_{\widetilde{\mathcal{P}}\left(s\vert d\right)} ,
\end{align}
where $\widetilde{\mathcal{P}}$ is a Gaussian probability distribution and $\widetilde{\mathcal{H}}$ is the respective Hamiltonian. $\left\langle ... \right\rangle_{\widetilde{\mathcal{P}}}$ indicates an averaging procedure of the quantity in the brackets with respect to $\widetilde{\mathcal{P}}$. Minimizing the above expression can be interpreted as a fit of $\widetilde{\mathcal{P}}$ to $\mathcal{P}$.
This approximated posterior is then easily evaluated via sampling due to its Gaussian nature. \\
Both the Hamiltonian and the KL functional can be conveniently implemented into NIFTy \citep{NIFTY_2013A&A...554A..26S, NIFTY_2017arXiv170801073S, NIFTY}, which in its newest version then takes care of calculating the correct gradients via auto-differentiation and provides a library of suitable minimization schemes for optimizing Eq. \eqref{eq:KL}. Convergence is assessed via predefined convergence criteria and the stability is tested via starting the inference from different (both deliberately and randomly chosen) initial conditions. For further details on the here used MGVI technique see \citet{JAKOB_2_2019arXiv190111033K}.
\end{appendix}

\end{document}